\documentclass[aps,prb,twocolumn,superscriptaddress]{revtex4-2}
\usepackage{graphicx}  
\usepackage{dcolumn}   
\usepackage{bm}        
\usepackage{amssymb}   
\usepackage{hyperref}
\usepackage{amsmath}
\DeclareMathOperator{\Tr}{Tr}
\usepackage{bbold}
\usepackage{xcolor}
\usepackage{footnote}
\usepackage{slashed}
\usepackage{physics}
\usepackage{float}
\usepackage{caption}
\usepackage{subcaption}
\usepackage{lipsum, babel}
\usepackage[toc,page]{appendix}

\begin{document}

\title{Anomalous Chiral Anomaly in Spin-1 Fermionic Systems}
\author{Shantonu Mukherjee}
\email{shantanumukherjeephy@gmail.com}
\affiliation{Department of Physics, Indian Institute of Technology Bombay, Powai, Mumbai 400076, India}
\author{Sayantan Sharma}
\email{sayantans@imsc.res.in}
\affiliation{The Institute of Mathematical Sciences, a CI of Homi Bhabha National Institute, Chennai 600113, India}
\author{Hridis K. Pal}
\email{hridis.pal@iitb.ac.in}
\affiliation{Department of Physics, Indian Institute of Technology Bombay, Powai, Mumbai 400076, India}

\date{\today}

\begin{abstract}
Chiral anomaly is a key feature of Lorentz-invariant quantum field theories: in presence of parallel external electric and magnetic fields, the number of massless Weyl fermions of a given chirality is not conserved. In condensed matter, emergent chiral fermions in Weyl semimetals exhibit the same anomaly, directly tied to the topological charge of the Weyl node, ensuring a quantized anomaly coefficient. However, many condensed matter systems break Lorentz symmetry while retaining topological nodes, raising the question of how chiral anomaly manifests in such settings. In this work, we investigate this question in spin-1 fermionic systems and show that the conventional anomaly equation is modified by an additional nontopological contribution, leading to a nonquantized anomaly coefficient. This surprising result arises because spin-1 fermions can be decomposed into  2-flavor Weyl fermions coupled to a Lorentz-breaking, momentum-dependent non-Abelian background potential. The interplay between this potential and external electromagnetic fields generates the extra term in the anomaly equation. Our framework naturally generalizes to other Lorentz-breaking systems beyond the spin-1 case.
\end{abstract}

\maketitle

\section{Introduction}\label{Sec-I}
Quantum anomalies refer to the violation of classical conservation laws, corresponding to some underlying symmetry, in the presence of quantum effects. They have been traditionally discussed in the context of high-energy physics, which is governed by Lorentz-invariant quantum field theories~\cite{bilal2008lectures,tHooft:1979rat}. A prominent example is the chiral anomaly, which refers to the non-conservation of the particle number of a specific chirality due to quantum effects~\cite{Adler:1969gk, Bell:1969ts,PhysRevLett.42.1195}. In its simplest form, it appears in free massless Weyl fermions in the presence of external  electromagnetic fields, and is expressed as ($\hbar=1$)~\cite{Peskin:1995ev}: 
\begin{equation}\label{anomalous non.cons.}
\partial_\mu J^{5\mu}- n\frac{e^2}{2\pi^2} \vec{E}\cdot \vec{B}=0,
\end{equation}
where $J^{5\mu}$ is the axial current, $\vec{E}$ and $\vec{B}$ are external electric and magnetic fields, and $n$ is the number of flavors of Weyl fermions.

In recent years, condensed matter systems have emerged as an exciting platform to study chiral anomaly~\cite{NIELSEN1983389, PhysRevB.86.115133, PhysRevLett.111.027201,PhysRevB.88.104412, PhysRevLett.113.247203, li2016chiral,ong2021experimental, PhysRevB.89.085126, 1997Natur.386..689B}. While Lorentz symmetry is absent at a fundamental level in crystalline solids, it can emerge effectively at low energies in materials such as Weyl semimetals. These systems host low-energy quasiparticles governed by the same effective field theories that describe relativistic Weyl fermions in high-energy physics~\cite{RevModPhys.90.015001, 10.1093/acprof:oso/9780199564842.001.0001}. This connection has enabled experimental studies of anomaly-induced transport phenomena in accessible table-top setups~\cite{ong2021experimental}. Weyl semimetals are characterized by an even number of band-touching points, or Weyl nodes, in three-dimensional momentum space. These nodes act as sources or sinks of Berry curvature, analogous to magnetic monopoles in momentum space, and are associated with a topological charge of $\pm 1$. For n-flavored Weyl systems, the topological charge thus becomes $\pm n$. This is the same integer that appears in Eq. (\ref{anomalous non.cons.}), allowing it to acquire a topological interpretation, with $n$ serving as a topological invariant.

Topology is not inherently tied to Lorentz invariance; it is possible to have  Lorentz symmetry broken at low energies, yet the topological charge remains nontrivial. This raises a fundamental question: what is the fate of chiral anomaly in such systems? Surprisingly, despite the prevalence of such systems in condensed matter physics, this question has attracted significantly less attention compared to the Lorentz-invariant Weyl semimetal case. The few studies that have addressed this, such as those focusing on multi-Weyl semimetals where the low-energy bands feature nonlinear dispersion with momentum, suggest that the conventional anomaly equation, Eq. (\ref{anomalous non.cons.}), remains valid~\cite{PhysRevB.96.085201,Lepori:2018vwg,PhysRevResearch.2.013007,PhysRevD.101.045007,gao2022chiral}.

In this paper, we explore chiral anomaly in systems with broken Lorentz invariance using the model of a spin-1 fermionic system. These systems, unlike multi-Weyl semimetals, are characterized by three bands touching at a point, comprising two linearly dispersing bands and one flat band~\cite{bradlyn2016beyond,PhysRevLett.119.206402}. Our main finding is that such a system admits a generalized chiral anomaly equation that renders Eq.~(\ref{anomalous non.cons.}) invalid. Specifically, in addition to the topological contribution, there is an additional contribution--an anomalous term which is non-topological. This arises due to the fact that the effective field theory of spin-1 fermions can be expressed as $2$-flavor Weyl fermions coupled to a background  non-Abelian potential that is \emph{momentum dependent}. In the presence of external electromangetic fields, this leads to an additional effective coupling between the electromagnetic potential and the non-Abelian potential, giving rise to the extra contribution in the anomaly equation.

The manuscript is organized as follows. In Sec-\ref{Sec-II} we introduce the model and construct an effective formulation of the problem. In Sec-\ref{Sec-III} we derive the anomaly equation for these systems and show how the anomaly equation is modified. In Sec-\ref{Sec-IV} we discuss possible observational consequences of the modified anomaly equation. Finally, in Sec~\ref{Sec-V}, we summarize our results and provide a general discussion of future directions and outlook.

\begin{figure*}[]
\centering
\includegraphics[width=1.0\textwidth]{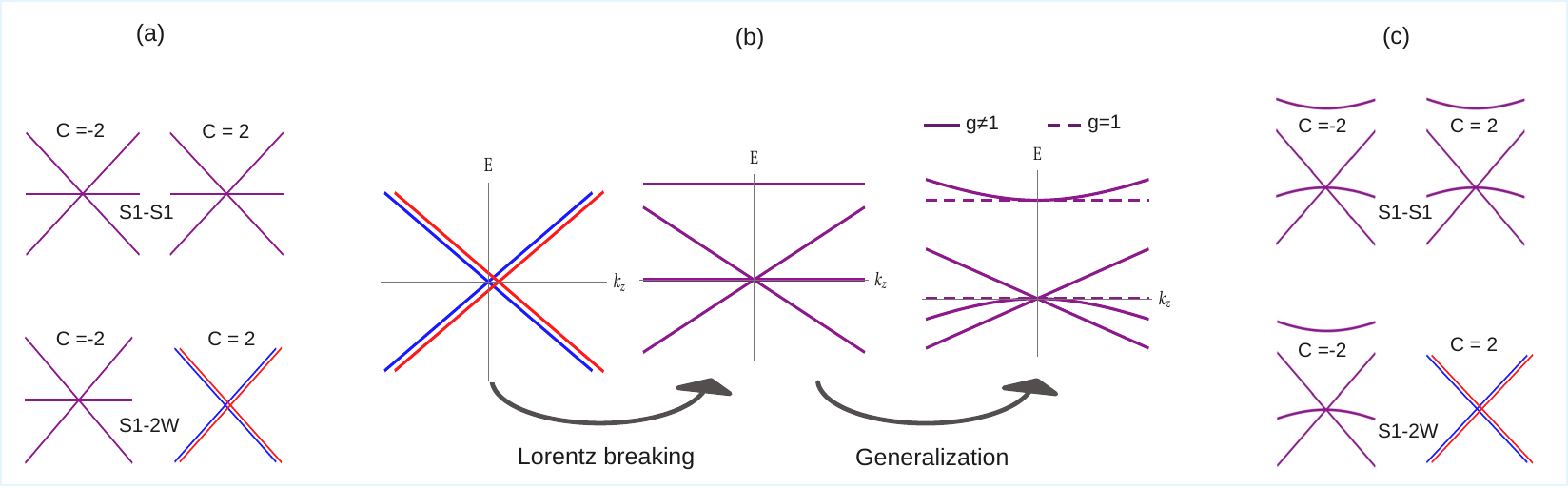}
\caption{(a) Two Lorentz-violating scenarios: the $S1$–$S1$ model, consisting of two spin-1 nodes with opposite topological charges $C$, where Lorentz symmetry is broken at both nodes; and the $S1$–$2W$ model, consisting of a spin-1 node and a double-Weyl node with opposite charges, where only the spin-1 node breaks Lorentz symmetry. (b) Structure of a spin-1 node: Three spin-1 bands, together with a higher-energy flat band, emerge from two-flavor Weyl fermions via band mixing and a Lorentz-violating perturbation [Eq.~(\ref{non-Abelian3})]. A tunable parameter $g$ generalizes this setup: the two flat bands at $g=1$ (dashed) acquire dispersion at large momentum when $g\ne 1$ (solid), but the topological charge and the linearly dispersing bands are unaffected by $g$ [Eq.~(\ref{non-Abelian4})]. (c) Final formulation: The generalized models studied for chiral anomaly, corresponding to the $S1$–$S1$ and $S1$–$2W$  cases shown in (a) [Eqs.~(\ref{eff hamiltonian}) and (\ref{effs12w}), respectively]. }
\label{fig1}
\end{figure*}

\section{Model}\label{Sec-II}
We consider a fermionic system described by the Hamiltonian $H= \vec{S}\cdot \vec{k}$ at some point in the $k-$space, where $\vec{S}$ are generators of $SO(3)$ group which serves as spin matrices for spin-$1$ particles (spin in this work implies pseudospin and not the actual spin). The explicit form of the Hamiltonian is:
\begin{equation}
   \tilde H_{\text{S1}}= v\begin{pmatrix}
        k_z & \frac{(k_x-ik_y)}{\sqrt{2}} & 0 \\
        \frac{(k_x+ik_y)}{\sqrt{2}}& 0 & \frac{(k_x-ik_y)}{\sqrt{2}}\\
        0& \frac{(k_x+ik_y)}{\sqrt{2}} & -k_z 
    \end{pmatrix}, 
    \label{Hspin1}
\end{equation}
where $v$ is the Fermi velocity; hereafter, we put $v=1$. The corresponding energy spectrum comprises two dispersive bands of energy $\pm k$ and a flat band at zero energy. The dispersive bands are topological while the flat band is nontopolgical, giving the node a topological charge of $\pm 2$. The Hamiltonian is inspired by the CoSi-family of compounds where such a scenario naturally arises at the $\Gamma$ point of the Brillouin zone~\cite{bradlyn2016beyond,PhysRevLett.119.206402, 2019PhRvL.122g6402T}.

By the Nielsen--Ninomiya theorem~\cite{Nielsen:1981hk}, the total Berry monopole charge in the Brillouin zone must vanish. As a minimal model, we consider a second spin-1 node of opposite topological charge $\mp2$, and define the resulting system as an $S1-S1$ system [ Fig.~\ref{fig1} (a)]. However, this symmetric setup is not the only possibility. In the CoSi-family of compounds, for instance, the compensating node at the \( R \) point hosts a pair of doubly-degenerate Weyl cones instead. We study this scenario as well, referring to this more asymmetric configuration as the $S1-2W$ system  [Fig.~\ref{fig1} (a)] \footnote{In reality, the situation is more complex: the two nodes lie at different energies and spin degeneracy is lifted by spin--orbit coupling. We neglect these effects here for simplicity.}. Our motivation for studying these two models is as follows. The $S1-S1$ model provides a natural setting to explore the effects of Lorentz symmetry breaking--intrinsic to spin-1 fermions--on chiral anomaly, which is the central focus of this work. The $S1$–$2W$ model, on the other hand, allows us to additionally explore the impact of asymmetry between nodes of opposite topological charge on the anomaly—-a question that remains unexplored to the best of our knowledge. In both cases, the Fermi energy is assumed to lie at zero energy (node).

Chirality is a well-defined quantum number only for massless particles within a Lorentz-invariant framework. Therefore, the first step in deriving the chiral anomaly in either the $S1-S1$ or the $S1-2W$ system is to decompose the Hamiltonian of a spin-1 node into a Lorentz-invariant part and a Lorentz-symmetry-breaking term. The key to this decomposition comes from the insights provided by the $S1-2W$ scenario in Fig.~\ref{fig1}(a). Specifically, the states corresponding to the low-energy doubly-degenerate Weyl bands at the R-point can be interpreted as a two-flavor Weyl field. As the four bands evolve toward the $\Gamma$-point through higher-energy sectors, they hybridize, resulting in three bands that touch at a single point while one band becomes gapped. This suggests that the three spin-$1$ bands, together with the gapped band, can be understood as arising from the two-flavor Weyl field that combines to form triplet and singlet states, whose degeneracy is broken due to the effect of a Lorentz symmetry-breaking term, as depicted in Fig.~\ref{fig1}(b). The Hamiltonian in these two basis, represented as $\frac{1}{2}\otimes\frac{1}{2}$ and $1\oplus 0$, respectively, are related to each other by an unitary transformation with a matrix $U$. Thus, we can write
\begin{eqnarray}
        \tilde H^{\lambda}_{\text{S1}}&=& U^{\dagger}\left(H_{\text{2W}}+h_{\text{LB}}^\lambda\right)U\nonumber\\
        &\equiv&\tilde H_{\text{2W}}+\tilde h_{\text{LB}}^\lambda,
        \label{Hlambdaspin1}
\end{eqnarray}
where $\tilde H^{\lambda}_{\text{S1}}\equiv \begin{pmatrix}
    \tilde H_{\text{S1}} && 0\\
    0 && \lambda
\end{pmatrix}$,
 with $\lambda$ denoting the energy difference between the gapped and ungapped bands at $k=0$. The Hamiltonian $H_{\text{2W}}= \Vec{\Sigma}\cdot\Vec{k}$ describes the $2$-flavor Weyl system, with $\Vec{\Sigma}=\Vec{\sigma}\otimes \mathbb{1}$ in the spin-flavor space. The term $h_{\text{LB}}^\lambda$ breaks Lorentz symmetry explicitly.

The goal now is to derive $h_{\text{LB}}^\lambda$, which proceeds via three steps.  First, we identify $U$ and compute $\tilde H_{\text{2W}}$; next, we subtract $\tilde H_{\text{2W}}$ from $\tilde H^{\lambda}_{\text{S1}}$ to obtain $\tilde h_{\text{LB}}^\lambda$; and finally, we perform an inverse transform on $\tilde h_{\text{LB}}^\lambda$ to obtain $h_{\text{LB}}^\lambda$. The details of the derivation are provided in the Appendix~\ref{Appendix B}
Following this procedure we obtain,
\begin{equation}
    h_{\text{LB}}^\lambda = \lambda \kappa_\lambda + k_i\kappa_i,\quad i=1,2,3
\end{equation}
where the matrices \( \kappa_\lambda \) and \( \kappa_i \) are
\begin{equation}
    \kappa_i = \frac{1}{2} ( \sigma_0 \otimes s_i - \sigma_i \otimes s_0),  
    \quad \kappa_\lambda = \frac{1}{4} (\sigma_0 \otimes s_0 - \sigma_i \otimes s_i),
\end{equation}
where \( \sigma_0 \) and \( s_0 \) are $2\times 2$ identity matrices. This suggests an underlying \( U(2) \) symmetry of the Hamiltonian, allowing us to interpret \( s_0, s_i \) as the generators of this symmetry group and the momentum components \( k_i \) as elements of an emergent non-Abelian potential. Consequently, we can rewrite the Lorentz-breaking term in a more compact form as 
\begin{equation}\label{non-Abelian2}
h_{\text{LB}}^\lambda= \mathcal{A}^a_\mu\, \sigma_\mu \otimes s_a,\,\, \{a,\mu\}=\{0,1,2,3\},
\end{equation} 
where the non-Abelian fields $A^a_\mu$ belong to a generalized non-Abelian \emph{color} group with components
\begin{equation}
\mathcal{A}^0_i= \frac{-k_i}{2},\, \mathcal{A}^i_0= \frac{k_i}{2},\,\mathcal{A}^0_0= \frac{\lambda}{4},\, \mathcal{A}^i_i= \frac{-\lambda}{4},\, \mathcal{A}^j_i=0.\label{non-Abelian-comp}
\end{equation}
In terms of the emergent non-Abelian fields, the Hamiltonian of spin-1 excitations (together with the gapped band) can be written as 
\begin{equation}\label{non-Abelian3}
 H^{\lambda}_{\text{S1}} = H_{\text{2W}}+h^{\lambda}_{\text{LB}}= H_{\text{2W}}+\mathcal{A}^a_\mu\, \sigma_\mu \otimes s_a,
\end{equation}
which completes the decomposition prescribed in Eq.~(\ref{Hlambdaspin1}). Note that $H^{\lambda}_{\text{S1}}=U\tilde H^{\lambda}_{\text{S1}}U^\dagger$. The above equation implies that the Lorentz breaking term can be viewed as a background non-Abelian potential.  
Importantly, this potential is momentum-dependent—-a feature that plays a crucial role in the physics that follows.

One final step remains before we embark on the anomaly calculation. Instead of working directly with Eq.~(\ref{non-Abelian3}), we consider a generalized form of the Hamiltonian:
\begin{eqnarray}
     H^{\lambda,g}_{\text{S1}}&=& H_{\text{2W}}+h^{\lambda,g}_{\text{LB}}= H_{\text{2W}}+\mathcal{A}^a_\mu(g) \sigma_\mu \otimes s_a,\label{non-Abelian4}\\
     \mathcal{A}^a_\mu(g)&=&\mathcal{A}^a_\mu\vert _{k_i\rightarrow gk_i},\label{gabelian}
\end{eqnarray}
where \( g \in \mathbb{R} \) is a tunable real parameter, and our original model is recovered when \( g = 1 \). The motivation for introducing this generalization will become clear shortly.  The eigenvalues of $H^{\lambda,g}_{\text{S1}}$ are given by $E = \pm k,\,\, \frac{1}{2} \left[\lambda \pm \sqrt{4(1-g)^2 k^2 + \lambda^2} \right]$. The two linearly dispersing bands remain unaltered. The two flat bands at $E=0,\lambda$ acquire a dispersion at large momentum $k\gtrsim \vert\frac{\lambda}{1-g}\vert$, below which they are essentially flat. This is depicted in Fig.~\ref{fig1}(b). Importantly, the topological structure remains unchanged: the two dispersing bands have the same Chern number $\pm 2$ as before while the remaining quasi-flat bands are non-topological; see Appendix~\ref{Appendix C} for more details~\footnotemark[\value{footnote}]. Thus, Eq.~(\ref{non-Abelian4}) defines a one-parameter family of Hamiltonians labeled by $g$, all of which reduce to the same spin-1 Hamiltonian in the small-momentum limit. The effect of $g$ is merely to introduce a dispersion in the non-topological flat bands at large momenta, leaving the topological linear bands unchanged. In real systems, this arises from lattice effects, beyond the continuum limit.

\section{Derivation of Chiral Anomaly}\label{Sec-III}
We now turn to the calculation of the chiral anomaly. We begin with the  \( S1-S1 \) system [Fig.~\ref{fig1}(c)]. 
The corresponding Hamiltonian can be written as
\begin{equation}\label{eff hamiltonian}
\mathcal{H}^{\lambda,g}_{\text{S1-S1}}=    \begin{pmatrix}
  H^{\lambda,g}_{\text{S1}}& 0\\
0& -H^{\lambda,g}_{\text{S1}} 
 \end{pmatrix}= \mathcal{H}_{\text{2W-2W}} + \mathcal{H}^{\lambda,g}_{\text{LB,S1-S1}},
\end{equation}
where the two terms in the Hamiltonian are 
\begin{equation}
    \mathcal{H}_{\text{2W-2W}}= \begin{pmatrix}
  H_{\text{2W}}& 0\\
0& -H_{\text{2W}} 
 \end{pmatrix},\,\, \mathcal{H}^{\lambda,g}_{\text{LB, S1-S1}}= \begin{pmatrix}
  h_{\text{LB}}^{\lambda,g}& 0\\
0& -h_{\text{LB}}^{\lambda,g} 
 \end{pmatrix}.
\end{equation}
We use the path integral technique due to Fujikawa~\cite{PhysRevLett.42.1195} to arrive at the anomaly equation. 
Coupling the system to external electromagnetic fields $A_\mu$ through a minimal coupling, $\partial_{\mu} \rightarrow \partial_\mu + ie A_\mu$, the low energy effective action of the system can be written in a covariant notation as (see Appendix~\ref{Appendix D} for more details)
\begin{equation}\label{eff action 2}
S^{\lambda,g}_{\text{S1-S1}}= \int\, d^4x\, ~\bar{\psi} (x)\left[ \slashed{D}+ \Gamma_0 \mathcal{H}_{\text{LB, S1-S1}}^{\lambda,g}\right]\psi(x)\,\ .
\end{equation}
The action is written in the Euclidean time direction obtained by performing a Wick rotation $\tau=it$. The Clifford algebra used in this work in the Euclidean signature is $\{\Gamma^\mu,\Gamma^\nu\} = - 2 \delta^{\mu\nu} \mathbb{1}.$
We emphasize that $\mathcal{H}_{\text{LB,S1-S1}}^{\lambda,g}$ also couples to the external electromagnetic field since it contains the derivative operator $\partial_\mu$.
Under local $U(1)$ and chiral transformations on the fermionic fields denoted as $\psi (x)\rightarrow e^{i\alpha}(x)\psi(x),\,\,\bar{\psi}(x) \rightarrow \bar{\psi} (x)e^{-i\alpha(x)}$ and $\,\,\psi(x)\rightarrow e^{i\alpha(x) \gamma_5\otimes \mathbf{1}}\psi(x),\,\, \bar{\psi}(x)\rightarrow \bar{\psi}(x)e^{i\alpha(x) \gamma_5\otimes \mathbf{1}}$, respectively, the corresponding \emph{effective} vector and axial currents, $J$ and $J^5$ respectively, are found to be
\begin{align}
&J^0=  \bar{\psi} \left[\gamma^0 \otimes \mathbb{1} \right] \psi,\\
& J^{i}= \bar{\psi} \left[\left(1-\frac{g}{2}\right)\gamma^i \otimes \mathbb{1}- \frac{i\,g}{2}\gamma^5\gamma^0\otimes s^i \right] \psi,\\
& J^{50}=  \bar{\psi} \left[\gamma^0 \gamma^5\otimes \mathbb{1} \right] \psi,\\
&J^{5i}= \bar{\psi} \left[\left(1-\frac{g}{2}\right)\gamma^i \gamma^5\otimes \mathbb{1}+  \frac{i\,g}{2}\gamma^0\otimes s^i \right] \psi.
\end{align}
The spatial and temporal components of the currents are not related to each other as observed in relativistic quantum field theories as a consequence of the broken Lorentz invariance. Both these currents can be shown to be conserved classically.
However the expectation value of the effective chiral current is not conserved when quantum effects are included. Within the path integral formalism, the non-conservation of the chiral current can be understood from the fact that the path integral measure of fermion fields is not invariant under chiral transformation~\cite{PhysRevLett.42.1195,Fujikawa:2004cx}. 
Computing this for our model, we find the following change in the action (see Appendix~\ref{Appendix F} for details):
\begin{eqnarray}
    \label{CA action gen}
\delta S^{\lambda, g}_{\text{S1-S1}}&=& -\int\,d^4x\,i\,\alpha(x)\bigg(\partial_\mu J^5_\mu+\bigg[\frac{2e^2}{16\pi^2} \varepsilon^{\mu\nu\rho\lambda}F^{\mu\nu}F^{\rho\lambda}\,\, \nonumber\\
&+&  \frac{4e}{32\pi^2} \ \text{Tr}[\gamma^5 \gamma_\mu\gamma_\nu\gamma_\alpha\gamma_i] F_{\mu\nu} [D_\alpha, \mathcal{A}_i^0(g)] \bigg] \bigg).
\end{eqnarray}
The first term within the square brackets in Eq.~\eqref{CA action gen} is the familiar contribution for the 2-flavor Weyl fermions while the second term is an additional contribution that arises entirely due to the Lorentz breaking component in the Hamiltonian.  Equation (\ref{CA action gen}) leads to the following anomaly equation (see Appendix~\ref{Appendix F} for details):
\begin{equation}
\partial_\mu J^{5\mu}- \left(2-3g \right) \frac{e^2}{2\pi^2} \vec{E}\cdot \vec{B}=0.
\label{result1}
\end{equation}
This result is distinctly different from  Eq.~(\ref{anomalous non.cons.}) and is the first important finding of this Letter.  The anomaly coefficient $n$ in Eq.~(\ref{anomalous non.cons.}) is replaced by $n'=2-3g\equiv n-3g$ in Eq.~(\ref{result1}), which is no longer quantized since $g$ can take any real value. Recall that $g$ in Eq.~(\ref{non-Abelian4}) affects only the non-topological flat bands, leaving the topological linear bands unaffected. Therefore, the additional $g$-dependent term in the anomaly reflects a non-topological origin. 
When $g=1$, corresponding to our original spin-1 Hamiltonian in Eq.~(\ref{Hspin1}), $|n'|=1$ instead of $2$ as one would expect based on conventional understanding (see also~\cite{Lepori:2018vwg}). Although quantized in this case, the result arises from a fortuitous cancellation between topological and non-topological contributions.  Intriguingly, when $g=2/3$, this cancellation is exact and the chiral current remains conserved! 

The essential observation is that field theories with broken Lorentz invariance can exhibit chiral anomalies in a qualitatively different way from their Lorentz-invariant counterparts: topology alone no longer dictates anomaly and quantization of the coeffecient is no longer guaranteed. While Lorentz symmetry breaking is necessary for this deviation, it is not sufficient. The additional non-topological contribution appears only when the commutator \([D_\alpha, \mathcal{A}_i^0(g)] \neq 0\), which, in our case, is ensured by the momentum dependence of the non-Abelian fields in Eqs.~(\ref{non-Abelian4}) and~(\ref{non-Abelian-comp}). This stands in contrast to multi-Weyl semimetals investigated earlier, which also lack Lorentz symmetry but exhibit quantized anomalies~\cite{PhysRevResearch.2.013007}. In those systems, the effective non-Abelian potentials are momentum-independent, hence \( [D_\alpha, \mathcal{A}_i^0] = 0 \), and the anomaly reduces to its conventional topological form (see Appendix~\ref{Appendix H}] for details). In our model as well, setting $g=0$ removes the momentum-dependence of $\mathcal{A}$, ~cf. Eq.~\ref{gabelian}], and the anomaly coefficient in Eq.~\ref{result1} reverts to its standard quantized value.

We next turn to the $S1-2W$ model which involves two inequivalent chiral nodes [Fig.~\ref{fig1}(c)]. In this case, the effective non-Abelian potential couples to fermions at one node while the other node remains unaffected, resulting in an effective action (see Appendix~\ref{Appendix E}),
\begin{equation}\label{eff action in equiv}
S^{\lambda,g}_{\text{S1-2W}}= \int\, d^4x\, \bar{\psi}\slashed{D}^{\lambda,g}_{\text{S1-2W}}\psi,\,\,\slashed{D}^{\lambda,g}_{\text{S1-2W}}= \slashed{D}+  \Gamma_0 \mathcal{H}^{\lambda,g}_{\text{LB},\,\text{S1-2W}},
\end{equation}
where $H^{\lambda,g}_{\text{LB},\,\text{S1-2W}}$ is defined as
\begin{equation}
\mathcal{H}^{\lambda,g}_{\text{LB},\,\text{S1-2W}}= \begin{pmatrix}
 0 & 0\\
0 & -h^{\lambda,g}_{\text{LB}}
 \end{pmatrix}.\label{effs12w}
\end{equation}
The calculation of chiral anomaly follows along similar lines as in the $S1-S1$ model, except for one difference which requires special attention: the modified Dirac operator $\slashed{D}^{\lambda,g}_{\text{S1-2W}}$ is now non-Hermitian. The procedure is detailed in the Appendix~\ref{Appendix F}. Finally, we obtain 
 \begin{equation}\label{Ac with gen pert}
\partial_\mu J^{5\mu}-\left( 2-\frac{3g}{2}\right)\frac{e^2}{2\pi^2} \vec{E}\cdot \vec{B} =0.
\end{equation}
Once again, we observe that the anomaly term acquires a $g-$dependent anomalous contribution resulting from the non-Abelian term in the Hamiltonian. Notably, for $g = 1$, we obtain $n' = 1/2$ instead of 1, indicating a fractional contribution. This can be attributed to the asymmetric coupling of the effective non-Abelian potential at the two nodal points. This, in turn, provides a quantitative answer to how nodal asymmetry in band structure influences chiral anomaly. This constitutes the second key finding of our work.

\section{Observable Consequence}\label{Sec-IV}
\begin{figure}[h]
\includegraphics[width=0.9\linewidth]{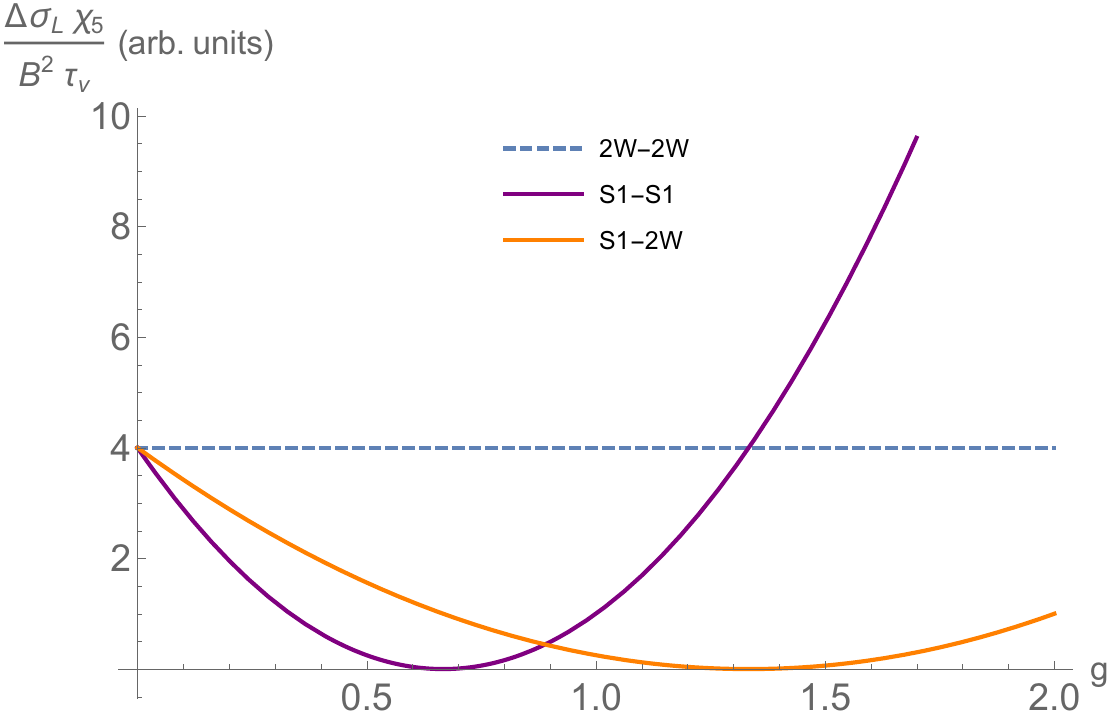}
\caption{%
Anomaly--induced longitudinal magnetoconductivity $\Delta\sigma_L\propto n'^2$ as a function of the non--topological coupling $g$.
The $S1-S1$ and $S1-2W$ cases both show dependence on $g$ arising from broken Lorentz symmetry, unlike the Lorentz symmetric $2W-2W$ case, shown in dashed for comparison.}
\label{fig:sigma_vs_g}
\end{figure}
The modified anomaly equation implies a corresponding revision of the anomaly--related observables previously discussed for Weyl systems. 
As an illustration, we consider the longitudinal magnetoconductivity (LMC): a positive LMC (i.e., negative longitudinal magnetoresistance) is a key experimental signature of chiral anomaly. 
Irrespective of microscopic details, the anomaly-induced contribution to the LMC corresponding to Eq.~(\ref{anomalous non.cons.}) can be written as ( see Appendix~\ref{Appendix I} for details):
\begin{equation}
    \Delta\sigma_{L} 
    = \frac{e^{4}}{4\pi^{4}}\,
      n^{2}\,
      \frac{\tau_{v}}{\chi_{5}}\,
      B^{2}.
\end{equation}
Here, $\tau_{v}$ is the relaxation time between the two chiral sectors and $\chi_{5}$ denotes the axial compressibility (or axial density of states). 
The coefficient of the $B^{2}$ term, apart from universal constants, contains two distinct ingredients: 
(i) the anomaly coefficient $n$, which originates from the chiral anomaly itself, and 
(ii) the material--dependent factors $\tau_{v}$ and $\chi_{5}$, which depend on the band structure and scattering mechanisms. 
For a Weyl node, $n=1$ and $\chi_{5}=\mu^{2}/(\pi^{2}v_{F}^{3})$, recovering the standard expression of Ref.~\cite{PhysRevB.88.104412}. 
This expression formally diverges as $\mu \rightarrow 0$, reflecting the vanishing density of states near the nodal point; in practice, this is regularized by an effective scale $\Gamma$ arising from disorder or interactions. 

In the spin-1 case, both factors differ. The anomaly coefficient is replaced by the non-quantized value $n'$, and the flat band ensures a finite axial compressibility $\chi_{5} \sim W_{f}/(\pi\Gamma_{f})$, where $W_{f}$ is the spectral weight associated with the flat band and $\Gamma_{f}$ its small broadening due to disorder. 
Consequently, the spin-1 system exhibits a finite longitudinal magnetoconductivity even at charge neutrality, with its magnitude varying continuously with the coupling $g$. This is demonstrated in Fig.~\ref{fig:sigma_vs_g} for both $S1-S1$ and $S1-2W$. A pronounced deviation from the Lorentz-symmetric limit highlights the non-topological contribution, providing a direct experimental handle on the underlying anomaly modification.\\

\section{Summary and Discussion}\label{Sec-V}
In this work, we studied chiral anomaly in fermionic systems that lack emergent Lorentz symmetry at low energies, focusing on spin-1 band crossings. We showed that the spin-1 Hamiltonian can be reformulated as an effective two-flavor Weyl theory coupled to a momentum-dependent non-Abelian background potential, allowing a unified quantum field–theoretic treatment of Lorentz-invariant and Lorentz-breaking effects. Using the Fujikawa method, we demonstrated that the chiral anomaly acquires an additional non-topological contribution originating from the momentum dependence of this potential. As a result, the anomaly coefficient is no longer fixed solely by the topological charge of the linear bands, in sharp contrast to conventional wisdom. We considered two classes of spin-1 systems: in the symmetric S1–S1 case, where both nodes break Lorentz symmetry, the anomaly is modified purely by Lorentz-breaking effects, while in the asymmetric S1–2W case, both nodal asymmetry and Lorentz symmetry breaking jointly modify the anomaly.

Our work opens several avenues for extension. 
Methodologically, the decomposition of a Lorentz--noninvariant system with topological charge $n$ into an $n$--flavor Lorentz--invariant Weyl term plus a non--Abelian potential is broadly applicable and can be extended to other condensed--matter settings and to anomalous symmetries beyond chirality. 
Beyond the longitudinal magnetoconductivity discussed here, other observables should be explored to identify further experimental manifestations of the anomalous chiral anomaly reported in this work. 

It is also intriguing that our results parallel certain non--Abelian anomaly phenomena in QCD. 
In vector--like gauge theories with massless quarks, the effective action resembles Eq.~\ref{eff action 2}, and non--topological configurations of parallel color electric and magnetic fields contribute to the anomaly~\cite{Lappi:2006fp,Mace:2016svc,Kharzeev:2001ev,Schlichting:2022fjc}. 
Similarly, the $g$--dependent anomalous term in Eq.~\ref{eff action in equiv} mirrors that of a single quark flavor in the electroweak theory, where only left--handed quarks couple to SU(2) gauge fields---the anomaly being exactly half of that in the vector--like case~\cite{Fujikawa:2004cx}. 
Further exploration of these analogies between condensed--matter and high--energy anomaly mechanisms remains an exciting direction for future research.

\begin{acknowledgments}
    S.M. would like to acknowledge the financial support from Indian Institute of Technology Bombay through Institute Post Doctoral Fellowship. H.K.P. would like to thank DST SERB, India for financial support via Grant No. CRG/2021/005453 and B. Roy for helpful discussions. 
\end{acknowledgments}

\begin{widetext}
    \begin{appendix}
        \section{Definition for $\Gamma$ matrices}\label{Appendix A}
In the main text as well as in the Appendices we have used the following notation for Dirac gamma matrices
\begin{align*}
    \Gamma_\mu= \gamma_\mu \otimes \mathbb{1},\,\,\,\, \Gamma^5= \gamma^5\otimes \mathbb{1},
\end{align*}
where $\gamma_\mu,\, \gamma^5$ are $4\times 4$ standard gamma matrices in chiral representation.

    \section{Spin-$1$ Hamiltonian from 2-Flavor Spin -$1/2$ Hamiltonian }\label{Appendix B}

Here in this section we shall try to write down detailed steps of derivation that would help the reader to follow the content of the main paper.  We first write down the Hamiltonian of a spin-$1$ and 2-flavor spin- $1/2$ system below

\begin{equation}\label{spin-1 matrices}
\tilde{H}_{\text{S1}}= \vec{S}\cdot \vec{k},\,\,
\begin{aligned}
& S_x =  \frac{1}{\sqrt{2}}\begin{pmatrix}
0 & 1 & 0\\
1 & 0 & 1\\
0 & 1 & 0
\end{pmatrix},\,
S_y = \frac{1}{\sqrt{2}} \begin{pmatrix}
0 & -i & 0\\
i & 0 & -i\\
0 & i & 0
\end{pmatrix},\,
 S_z =  \begin{pmatrix}
1 & 0 & 0\\
0 & 0 & 0\\
0 & 0 & -1
\end{pmatrix}.
\end{aligned}
\end{equation}
and 

\begin{equation}
H_{2W}= \vec{\Sigma}\cdot\vec{k},\,\,  \vec{\Sigma}= \vec{\sigma}\otimes \mathbb{1}.
\end{equation}
where $\sigma^i$ are the Pauli matrices which are (pseudo) spin matrices and $\mathbb{1}$ represents the operation on the flavor space. Below we illustrate on the argument given in the main paper that the states of a $2$-flavor Weyl fermions hybridize and form a spin-$1$ band crossing and a high energy band. To proceed, we notice that $2$-flavor Weyl Hamiltonian acts on the following Basis 
\begin{equation}\label{spin-flavor states}
    \ket{\psi} = \begin{pmatrix}
        \ket{\phi_A}\\ 
        \ket{\phi_B}
    \end{pmatrix} \otimes \begin{pmatrix}
        \ket{\chi_1}\\ 
        \ket{\chi_2}
    \end{pmatrix}.
\end{equation}
Out of these four basis states one can now form another set of ortho-normal basis states which transform as triplet and singlet states
\begin{align}
 & \text{Triplet:} \quad \ket{\phi_A}\otimes \ket{\chi_1}, \, \frac{1}{\sqrt{2}}\left(\ket{\phi_A}\otimes \ket{\chi_2} + \ket{\phi_B}\otimes \ket{\chi_1}\right),\, \ket{\phi_B}\otimes \ket{\chi_2}.\\
&  \text{Singlet:}\quad\frac{1}{\sqrt{2}}\left(\ket{\phi_A}\otimes \ket{\chi_2} - \ket{\phi_B}\otimes \ket{\chi_1}\right).
\end{align}
We note that if we interchange the spin indices $A,\, B$ and flavor indices $1,\, 2$ within each other the first three states transform symmetrically into each other forming a sub-space. Thus they form a triplet, while the fourth state transform anti-symmetrically forming a singlet. One can now find out a unitary transformation between these two sets of ortho-normal vectors in the following way. We represent the states in Eq~\eqref{spin-flavor states} as
\begin{equation}\label{spin-flavor rep}
    \ket{\phi_A}\otimes \ket{\chi_1}= \begin{pmatrix}
         1\\0\\0\\0
     \end{pmatrix},\,  \ket{\phi_A}\otimes \ket{\chi_2}= \begin{pmatrix}
         0\\1\\0\\0
     \end{pmatrix}\, \ket{\phi_B}\otimes \ket{\chi_1}=\begin{pmatrix}
         0\\0\\1\\0
     \end{pmatrix},\, \ket{\phi_B}\otimes \ket{\chi_2}= \begin{pmatrix}
         0\\0\\0\\1
     \end{pmatrix}.
\end{equation}
Then the triplet and singlet states can be written as 
\begin{equation}\label{triplet-singlet rep}
\begin{aligned}
  \tilde{\ket{\psi}}:\quad &\ket{\phi_A}\otimes \ket{\chi_1}=\begin{pmatrix}
         1\\0\\0\\0
     \end{pmatrix},\,\, \frac{1}{\sqrt{2}}\left(\ket{\phi_A}\otimes \ket{\chi_2} + \ket{\phi_B}\otimes \ket{\chi_1}\right)=   \frac{1}{\sqrt{2}}\begin{pmatrix}
         0\\1\\1\\0
     \end{pmatrix},\,\, \ket{\phi_B}\otimes \ket{\chi_2}= \begin{pmatrix}
         0\\0\\0\\1
     \end{pmatrix},\\
  &  \frac{1}{\sqrt{2}}\left(\ket{\phi_A}\otimes \ket{\chi_2} - \ket{\phi_B}\otimes \ket{\chi_1}\right) =  \frac{1}{\sqrt{2}}   \begin{pmatrix}
         0\\1\\-1\\0
     \end{pmatrix}.
\end{aligned}
\end{equation}
It is now quite easy to understand that the first set of states in Eq~\eqref{spin-flavor rep} transforms into the states in Eq~\eqref{triplet-singlet rep} by the following unitary transformations
 \begin{equation}\label{unitary trans}
 \ket{\psi}= U \tilde{\ket{\psi}},\quad U= \begin{pmatrix}
        1 & 0 & 0 & 0\\
        0 & \frac{1}{\sqrt{2}} & 0 & \frac{1}{\sqrt{2}}\\
        0 & \frac{1}{\sqrt{2}} &0 & -\frac{1}{\sqrt{2}}\\
        0 & 0& 1 & 0
    \end{pmatrix}. 
 \end{equation}
We shall show below that the Hamiltonian of a spin-$1$ particle can be obtained from that of $2$-flavor spin-$1/2$ fermions using the same transformations. We use this transformation in the following way
\begin{equation}\label{DW Trans Ham}
\begin{aligned}
    H_{\text{2W}} \rightarrow \tilde{H}_{\text{2W}}= U^\dagger H_{2W} U & = \begin{pmatrix}
        k_z & \frac{(k_x-ik_y)}{\sqrt{2}} & 0 & -\frac{(k_x-ik_y)}{\sqrt{2}}\\
        \frac{(k_x+ik_y)}{\sqrt{2}}& 0 & \frac{(k_x-ik_y)}{\sqrt{2}}& k_z\\
        0& \frac{(k_x+ik_y)}{\sqrt{2}} & -k_z & \frac{(k_x+ik_y)}{\sqrt{2}}\\
        -\frac{(k_x+ik_y)}{\sqrt{2}}& k_z &  \frac{(k_x-ik_y)}{\sqrt{2}}& 0
    \end{pmatrix}.
    \end{aligned}
\end{equation}
We see that the elements in the first $3\times 3$ block in the Hamiltonian of Eq~\eqref{DW Trans Ham} looks like the Hamiltonian of a spin-$1$ system. However, the unitary transformation would not change the eigen spectrum of the $2$-flavor Weyl system. To disentangle the spin-$1$ part one need to gap out the singlet state from the triplet states. This gapped band along with the spin-$1$ band crossing comprising of the triplet bands would then represent the hybridized band structure at the $\Gamma$ point. This is achieved via adding a momentum dependent potential term to the $2$-flavor Weyl Hamiltonian in Eq~\eqref{DW Trans Ham}. This is illustrated through the following decomposition
\begin{equation}\label{spin 1 in rotated basis}
 \tilde{H}^{\lambda}_{\text{S1}}=\begin{pmatrix}
        k_z & \frac{(k_x-ik_y)}{\sqrt{2}} & 0 & 0\\
        \frac{(k_x+ik_y)}{\sqrt{2}}& 0 & \frac{(k_x-ik_y)}{\sqrt{2}}& 0\\
        0& \frac{(k_x+ik_y)}{\sqrt{2}} & -k_z &0\\
       0& 0 & 0& \lambda
    \end{pmatrix} = \tilde{H}_{\text{2W}}+  \tilde{h}^{\lambda}_{\text{LB}},\quad  \tilde{h}^{\lambda}_{\text{LB}}=\begin{pmatrix}
       0 &0 & 0 & \frac{(k_x-ik_y)}{\sqrt{2}}\\
       0& 0 &0& -k_z\\
        0& 0 & 0 & -\frac{(k_x+ik_y)}{\sqrt{2}}\\
        \frac{(k_x+ik_y)}{\sqrt{2}}& -k_z &  -\frac{(k_x-ik_y)}{\sqrt{2}}& \lambda
    \end{pmatrix},
\end{equation}
where the Hamiltonian $\tilde{H}^{\lambda}_{\text{S1}}$ represents the spin-$1$ system written in the triplet basis along with the singlet band at some high energy $E=\lambda$. Thus spin-$1$ fermions can be formulated as $2$-flavor Weyl fermions subject to an internal Lorentz symmetry breaking potential which is represented by the second term in Eq~\eqref{spin 1 in rotated basis} written in the rotated basis. However, one can formulate a more generalized Hamiltonian by scaling the momentum $k \rightarrow g\, k$ such that the eigen spectrum and topological charges of the bands of the generalized Hamiltonian remains  same as that of the above Hamiltonian in Eq~\eqref{spin 1 in rotated basis}
\begin{equation}\label{Ham. gen. Trans}
\tilde{H}^{\lambda,g}_{\text{S1}}= \tilde{H}_{\text{2W}}+\tilde{h}^{\lambda, g}_{\text{LB}}(k),\quad \tilde{h}^{\lambda, g}_{\text{LB}}(k)= g \begin{pmatrix}
       0 &0 & 0 & \frac{(k_x-ik_y)}{\sqrt{2}}\\
       0& 0 &0& -k_z\\
        0& 0 & 0 & -\frac{(k_x+ik_y)}{\sqrt{2}}\\
        \frac{(k_x+ik_y)}{\sqrt{2}}& -k_z &  -\frac{(k_x-ik_y)}{\sqrt{2}}& \frac{\lambda}{g}
    \end{pmatrix}.
\end{equation}
In the original basis this generalized Lorentz symmetry breaking term looks like
\begin{equation}\label{explicit form tilde h_p}
h^{\lambda, g}_{\text{LB}}(k)= U\,\tilde{h}^{\lambda, g}_{\text{LB}}\, U^\dagger= g\,\begin{pmatrix}
0 & \frac{k^{-}}{\sqrt{2}} & - \frac{k^{-}}{\sqrt{2}} & 0\\
\frac{k^{+}}{\sqrt{2}} & - k_z + \frac{\lambda}{2g} & -\frac{\lambda}{2g}& -\frac{k^{-}}{\sqrt{2}}\\
-\frac{k^{+}}{\sqrt{2}} & -\frac{\lambda}{2g} &  k_z + \frac{\lambda}{2g} & \frac{k^{-}}{\sqrt{2}} \\
0 & -\frac{k^{+}}{\sqrt{2}} & \frac{k^{+}}{\sqrt{2}} & 0
\end{pmatrix}.
\end{equation}
However, as argued in our paper, this Lorentz symmetry breaking term can be formulated as a non-Abelian potential where the non-Abelian field components are momentum dependent. This can be shown explicitly by decomposing the matrix $h^{\lambda,g}_{\text{LB}}$ in terms of the momentum components as follows
\begin{equation}
h^{\lambda, g}_{\text{LB}} (k)= \lambda\kappa_\lambda + \kappa_i g k_i,\quad i=1,2,3.
\end{equation}
where the $\kappa$ matrices are given by
\begin{equation}
\label{kappa matrices}
\kappa_1= \frac{1}{2}\begin{pmatrix}
0&1&-1&0\\
1&0&0&-1\\
-1&0&0&1\\
0&-1&1&0
\end{pmatrix},\,\,
\kappa_2=\frac{i}{2} \begin{pmatrix}
0&-1&1&0\\
1&0&0&1\\
-1&0&0&-1\\
0&-1&1&0
\end{pmatrix},\,\,
\kappa_3= \begin{pmatrix}
0&0&0&0\\
0&-1&0&0\\
0&0&1&0\\
0&0&0&0
\end{pmatrix},\,\, \kappa_\lambda= \frac{1}{2}\begin{pmatrix}
0&0&0&0\\
0&1&-1&0\\
0&-1&1&0\\
0&0&0&0
\end{pmatrix}.
\end{equation}
These $\kappa$ matrices can be written in terms of Pauli matrices $\sigma_\mu,\, s_\mu$ in the following way
\begin{equation}
\kappa_i= \frac{1}{2}\left[\sigma_0 \otimes s_i- \sigma_i\otimes s_0 \right],\,\, \kappa_\lambda= \frac{1}{4}\left[\sigma_0\otimes s_0- \sigma_i\otimes s_i\right].
\end{equation} 
where $\sigma_0,\, s_0$ are two dimensional identity matrices. This decomposition of the Lorentz symmetry breaking term $h^{\lambda, g}_{\text{LB}}$ in terms of $s_0,\, s_i$, which are the generators of the $U(2)$ group, allows us to capture the underlying $U(2)$ symmetry and interpret corresponding momentum components as components of a non-Abelian $U(2)$ potential. This is expressed by the following equation
\begin{equation}\label{non-Abelian2}
h^{\lambda, g}_{\text{LB}}= \mathcal{A}^a_\mu(g)\, \sigma_\mu \otimes s_a,\,\, \{a,\mu\}=\{0,1,2,3\},
\end{equation} 
where the new tensor fields $\mathcal{A}^a_\mu$ are the component of a non-Abelian potential and are given by
\begin{equation}\label{non-Abelian fields}
\mathcal{A}^0_i= -g\frac{k_i}{2},\, \mathcal{A}^i_0= g\frac{k_i}{2},\,\mathcal{A}^0_0= \frac{\lambda}{4},\, \mathcal{A}^i_i= -\frac{\lambda}{4},\, \mathcal{A}^j_i=0~,~i,j=1,2,3.
\end{equation}
So the Hamiltonian of the spin-$1$ fermion can finally be written in terms of these non-Abelian fields as 
\begin{equation}\label{effective Spin $1$}
    H^{\lambda,g}_{\text{S1}}= H_{\text{2W}} +  \mathcal{A}^a_\mu(g)\, \sigma_\mu \otimes s_a.
\end{equation}
\section{Equivalence of eigen spectrum of $\tilde{H}_{\text{S1}}$ and $\Tilde{H}^{\lambda,g}_{\text{S1}}$}\label{Appendix C}

In this section we shall show that the original spin-$1$ fermion Hamiltonian in Eq~\eqref{spin-1 matrices} and that in Eq~\eqref{Ham. gen. Trans} are equivalent for different values of $g$ in the sense that in both cases the low energy eigen spectra are same and have the same topology. To demonstrate these points let us compare the cases-- one with $g=1$, which is by construction equivalent to the Hamiltonian of Eq~\eqref{spin-1 matrices} along with a flat band at energy $\lambda$, and another with a general value of $g$. To proceed we first calculate the normalized eigen states corresponding to the Hamiltonian in Eq~\eqref{Ham. gen. Trans} for $g=1$, which are 
\begin{align}
  &  \tilde{\psi}^+|_{g=1} = \frac{\sin^2\theta}{2(1+\cos\theta)}\left[\frac{(1+\cos{\theta})^2}{\sin^2{\theta}}e^{-i2\phi},\,\, \sqrt{2} \frac{(1+\cos{\theta})}{\sin{\theta}} e^{-i\phi},\,\, 1,\,\, 0 \right]^T,\\
  & \tilde{\psi}^-|_{g=1} =  \frac{\sin^2\theta}{2(1-\cos\theta)}\left[-\frac{(1-\cos{\theta})^2}{\sin^2{\theta}}e^{-i2\phi},\,\, -\sqrt{2} \frac{(1-\cos{\theta})}{\sin{\theta}} e^{-i\phi},\,\, 1,\,\, 0 \right]^T,\\
  &\tilde{\psi}^0|_{g=1}= \frac{1}{\sqrt{2} \csc{\theta}}\left[-e^{-i2\phi},\,\, \sqrt{2} \cot{\theta} e^{-i\phi},\,\, 1,\,\, 0 \right]^T,\\
  & \tilde{\psi}^\lambda|_{g=1}= [0,\, 0,\, 0,\, 1]^T.
\end{align}
These eigen states correspond to the eigen values $E= +k,-k,0,\lambda$ respectively. The topological charges of these bands are $n= +2, -2,0,0$ respectively. We observe that the first three states coincide with the eigen spectrum and topology of the eigen states of the Hamiltonian in Eq~\eqref{spin-1 matrices}. Let us now write down the eigen spectra for the effective Hamiltonian in Eq~\eqref{Ham. gen. Trans} for general g 
\begin{align}
   & \tilde{\psi}^+|_g = \frac{\sin^2\theta}{2(1+\cos\theta)}\left[\frac{(1+\cos{\theta})^2}{\sin^2{\theta}}e^{-i2\phi},\,\, \sqrt{2} \frac{(1+\cos{\theta})}{\sin{\theta}} e^{-i\phi},\,\, 1,\,\, 0 \right]^T,\\
  & \tilde{\psi}^-|_g = \frac{\sin^2\theta}{2(1-\cos\theta)}\left[-\frac{(1-\cos{\theta})^2}{\sin^2{\theta}}e^{-i2\phi},\,\, -\sqrt{2} \frac{(1-\cos{\theta})}{\sin{\theta}} e^{-i\phi},\,\, 1,\,\, 0 \right]^T,\\
   & \tilde{\psi}^0|_g= \frac{\xi_{-}(\lambda,g,k)}{\sqrt{\zeta_{-}^2(\lambda,g,k)+ \xi_{-}^2(\lambda,g,k) k^2 }} \left[-\frac{k \sin\theta}{\sqrt{2}}e^{-i\phi, },\, k \cos\theta, \frac{k \sin\theta}{\sqrt{2}}e^{i\phi}, \frac{\zeta_{-}(\lambda,g,k)}{\xi_{-}(\lambda,g,k)} \right]^T,\\
   & \tilde{\psi}^\lambda|_g= \frac{\xi_{+}(\lambda,g,k)}{\sqrt{\zeta_{+}^2(\lambda,g,k)+ \xi_{+}^2(\lambda,g,k) k^2 }} \left[-\frac{k \sin\theta}{\sqrt{2}}e^{-i\phi, },\, k \cos\theta, \frac{k \sin\theta}{\sqrt{2}}e^{i\phi}, \frac{\zeta_{+}(\lambda,g,k)}{\xi_{+}(\lambda,g,k)} \right]^T,\\
   & \xi_{\pm}(\lambda,g,k)= \lambda(g^2-2g +2) \pm g(g-2)\sqrt{4(g-1)^2 k^2 + \lambda^2},\\
   & \zeta_{\pm}(\lambda,g,k)= (g-1)\left[2g(g-2) k^2 + \lambda \left(\lambda \pm \sqrt{4(g-1)^2 k^2 + \lambda^2}\right)\right].
\end{align}
The eigen values corresponding to those eigen states above are given by $E= \pm k, \frac{1}{2} \left[\lambda \mp \sqrt{4 (g-1)^2k^2 + \lambda^2} \right]$ respectively. One can immediately check that these eigen states $\tilde{\psi}^+|_g,\tilde{\psi}^-|_g$ and the corresponding eigen values matches with $\tilde{\psi}^+|_{g=1},\tilde{\psi}^-|_{g=1}$ exactly while the states $\tilde{\psi}^0 |_g,\tilde{\psi}^\lambda|_g$ seem to differ from $\tilde{\psi^0}|_{g=1},\tilde{\psi}^\lambda|_{g=1}$. However, in the low energy limit $k<< \lambda$ these are equivalent to each other. It can also be checked explicitly that the states $\tilde{\psi}^+|_g,\tilde{\psi}^-|_g$ have topological charges $\pm 2$ while that for $\tilde{\psi}^0 |_g,\tilde{\psi}^\lambda|_g$ is zero. Thus we have shown that for different values of $g$ the Hamiltonian in Eq~\eqref{Ham. gen. Trans} represent the spin-$1$ system.
\section{Effective theory for the $\text{S1}-\text{S1}$ system}\label{Appendix D}
We now construct the effective Lagrangian of an auxiliary system whose left and right chiral nodes are represented by the spin-$1$ fermions having opposite chirality and are expressed by the Hamiltonian of Eq~\eqref{effective Spin $1$} 
\begin{equation}
    \mathcal{L}^{\lambda,g}_{\text{S1-S1}}= i\psi^\dagger\partial_t\psi - \psi^\dagger \mathcal{H}^{\lambda,g}_{S1-S1} \psi,\quad \mathcal{H}^{\lambda,g}_{S1-S1}\equiv \begin{pmatrix}
          H^{\lambda,g}_{\text{S1}} & 0\\
       0 & - H^{\lambda,g}_{\text{S1}}
    \end{pmatrix}.
\end{equation}
In presence of external electromagnetic fields and substituting the interaction in terms of the \emph{effective} non-Abelian fields the Lagrangian in Euclidean metric signature becomes,
\begin{equation}
    \mathcal{L}^{\lambda,g}_{\text{S1-S1}}= \bar{\psi} \slashed{D} \psi -i \begin{pmatrix}
0 & \, \sigma_0\\
-  \sigma_0 & 0
\end{pmatrix}\otimes \mathcal{A}^a_0(g)\,s_a
-i \begin{pmatrix}
0 &  \sigma_i\\
-  \sigma_i & 0
\end{pmatrix}\otimes \mathcal{A}^a_i(g)\,s_a,
\end{equation}
where we have written $\slashed{D}\equiv \Gamma_\mu D_\mu $ is the covariant derivative in presence of external 
electromagnetic fields $\vec{E}, \vec{B}$. Recognizing that
\begin{equation}
\begin{pmatrix}
0 & \, \sigma_0\\
-  \sigma_0 & 0
\end{pmatrix} = -i \gamma^5 \gamma^0,\,\,  \begin{pmatrix}
0 &  \sigma_k\\
-  \sigma_k & 0
\end{pmatrix}= \gamma^k,~~k=1,2,3~.
\end{equation}
 We can write the final form of the effective Lagrangian as

\begin{equation}\label{Action_Elegant}
    \mathcal{L}^{\lambda,g}_{\text{S1-S1}}= \Bar{\psi} \left(\gamma_\mu\otimes \mathbb{1} D_\mu + \Gamma_0 \mathcal{H}^{\lambda,g}_{\text{LB,S1-S1}}\right)\psi,\quad \Gamma_0\mathcal{H}^{\lambda,g}_{\text{LB,S1-S1}}=-\left( \gamma^5 \gamma_0\otimes \mathcal{A}^a_0(g)\,s_a
+i\gamma_k\otimes \mathcal{A}^a_k(g)\,s_a\right).
\end{equation}
\section{Effective theory for the $\text{S1}-\text{2W}$ system}\label{Appendix E}

By generalizing the above construction one can similarly obtain the effective theory for a system where one of the nodes is described by spin-$1$ fermion while the other is described by $2$ flavor Weyl fermion. Thus we get the following Lagrangian of the above mentioned system
\begin{equation}
    \mathcal{L}^{\lambda,g}_{\text{S1-2W}}= i\psi^\dagger\partial_t\psi - \psi^\dagger \mathcal{H}^{\lambda,g}_{\text{S1-2W}} \psi,\quad \mathcal{H}^{\lambda,g}_{\text{S1-2W}}= \begin{pmatrix}
         H_{\text{2W}} & 0\\
       0 & - H^{\lambda,g}_{\text{S1}}
    \end{pmatrix}.
\end{equation}
The corresponding effective action for this system
\begin{equation}\label{eff action in equiv nodes}
S^{\lambda,g}_{\text{S1-2W}}= \int\, d^4x\, ~\bar{\psi}(x)\slashed{D}^{\lambda,g}_{\text{S1-2W}}\psi(x),\,\,\slashed{D}^{\lambda,g}_{\text{S1-2W}}= \slashed{D}+  \Gamma_0 \mathcal{H}^{\lambda,g}_{\text{LB},\,\text{S1-2W}}.
\end{equation}
where $H^{\lambda,g}_{\text{LB},\,\text{S1-2W}}$ in this case is defined as 
\begin{equation}
\mathcal{H}^{\lambda,g}_{\text{LB},\,\text{S1-2W}}= \begin{pmatrix}
 0 & 0\\
0 & -h^{\lambda,g}_{\text{LB}}
 \end{pmatrix}
\end{equation}



\section{ Calculation of chiral anomaly in the path integral formalism}\label{Appendix F}
\subsection*{Case 1 : S1-S1 System}
Here we shall follow Fujikawa's method to calculate chiral anomaly in the above mentioned cases of interest. To proceed, we first write down the generating functional with the action in Eq~\eqref{Action_Elegant}
\begin{equation}\label{gen. functional}
\mathcal{Z}= \int \mathcal{D}\psi \mathcal{D}\bar{\psi} \exp\left[-\int\, d^4x\,~ \bar{\psi}(x)\slashed{D}^{\lambda,g}_{\text{S1-S1}}\psi(x)\right],\quad \slashed{D}^{\lambda,g}_{\text{S1-S1}}= \slashed{D}+  \Gamma_0 \mathcal{H}^{\lambda,g}_{\text{LB},\,\text{S1-S1}}
\end{equation}
We shall now expand fermionic fields $\bar{\psi}$ and $\psi$ in terms of a complete set of basis vectors. For the auxiliary case in which both Weyl nodes are described by spin-$1$ fermions of opposite chirality, we shall use the eigen states of $\slashed{D}^{\lambda,g}_{\text{S1-S1}}$ operator as it is a Hermitian operator in the Euclidean metric signature. However, for multi-Weyl systems in general the modified Dirac operator may not be Hermitian. In such cases we shall consider the eigen states of the Laplacian operator $\slashed{D}^\dagger\,\slashed{D}$ and $\slashed{D}\slashed{D}^\dagger$ \cite{Fujikawa:2004cx, PhysRevResearch.4.L042004}. As we shall see shortly, this method will be useful for a more general class of multi-Weyl systems. We assume that the $\phi_m(x)$ are the eigen states of the modified Dirac operator with eigen values $\lambda_m$
\begin{equation}\label{eigen eq for mod. Dirac}
\slashed{D}^{\lambda,g}_{\text{S1-S1}}\phi_n(x)= \lambda_m \phi_m.
\end{equation} 
The Grassmann valued fermionic fields can be thus represented as 
\begin{equation}
\psi= \sum_m C_m \phi_m(x),\,\, \bar{\psi}= \sum_m \bar{C}_m\phi^\dagger_m,
\end{equation}
where the coefficients $C_m$ and $\bar{C}_m$ are Grassmann numbers. The functional measure $\mathcal{D}\psi \mathcal{D}\bar{\psi}$ can now be written as\cite{Peskin:1995ev}
\begin{equation}
\mathcal{D}\psi \mathcal{D}\bar{\psi}= \prod_m dC_m d\bar{C}_m.
\end{equation}
For the infinitesimal chiral transformation of the fermionic fields i.e $\delta\psi(x)\simeq (i\alpha(x) \gamma_5\otimes \mathbf{1})\,\psi(x),\,\, \delta \bar{\psi}(x)\simeq \bar{\psi}(x)\, ( i\alpha(x) \gamma_5\otimes \mathbf{1})$, the Grassmann valued coefficients transform as
\begin{equation}\label{trans. of fermions}
C^\prime_p= \sum_{m}(\mathbb{1}+ K)_{pm} C_m,\, \bar{C}^\prime_p= \sum_m \bar{C}_m (\mathbb{1}+ K)_{mp},
\end{equation}
where the matrix elements of $K$ are given by
\begin{equation}
K_{pq}= i\int\, d^4x\,\alpha(x)\, \phi_p^\dagger\, \Gamma^5\, \phi_q~,~\Gamma^5= \gamma_5\otimes \mathbf{1}~.
\end{equation}
Due to this the functional measure transform as 
\begin{equation}
 \prod_m dC_m d\bar{C}_m\rightarrow \mathcal{J}^{-2}  \prod_m dC_m d\bar{C}_m,
\end{equation}
where $\mathcal{J}$ is the Jacobian for the transformation in Eq~\eqref{trans. of fermions} and has the following form
\begin{equation}\label{Jacobian}
\mathcal{J}= \det[\mathbb{1}+ K]= e^{\Tr \ln [\mathbb{1}+ K]}.
\end{equation}
The quantum effects are encoded in this Jacobian factor. For infinitesimal chiral transformation i.e. $\alpha$ to be small we have
\begin{equation}\label{Jacobian div}
\ln \mathcal{J}= \Tr[ K- \frac{K^2}{2}+\cdots]\simeq i\int\,d^4x\,\, \alpha\,\phi_m^\dagger\Gamma^5\phi_m,
\end{equation}
where the index m in summed over and also considered only the first term of the expansion. Expressing the eigen states of the modified Dirac operator in position basis as $\phi_n(x)= \bra{x}\ket{\phi_n}$ one can see easily that the term in the RHS of Eq~\eqref{Jacobian div} is the difference of two divergences and therefore undefined. To obtain a physically meaningful result we need to regularize the integral in Eq~\eqref{Jacobian div}. For that purpose we use the regulator $e^{-\frac{\lambda^2_m}{M^2}}$ into Eq~\eqref{Jacobian div} and after performing the trace calculation we eventually set the parameter $M$ to be infinitely large, such that,
\begin{equation}
\begin{aligned}
\ln \mathcal{J} &\simeq \lim_{M\to\infty}\,i\,\int\,d^4x\,\, \alpha(x)\,\phi_m^\dagger\,\Gamma^5 e^{-\frac{\lambda^2_m}{M^2}}\,\phi_m\\
&= \lim_{M\to\infty}\,i\,\int\,d^4x\,\, \alpha(x)\,\Tr\left[\bra{x}\,\Gamma^5 e^{-\frac{-\left(\slashed{D}^{\lambda,g}_{\text{S1-S1}}\right)^2}{M^2}}\ket{x}\right],
\label{eqn:Jacobian}
\end{aligned}
\end{equation}
where in the last line we have used the Eq~\eqref{eigen eq for mod. Dirac} and the completeness relation for the eigen states $\phi_m$: $\sum_m \ket{\phi_m}\bra{\phi_m}=1$. Now we just have to evaluate the integral in the above equation. For that we note that in our case $(\slashed{D}^{\lambda,g}_{\text{S1-S1}})^2$ is given by the following expression
\begin{equation}
\begin{aligned}
\left(\slashed{D}^{\lambda,g}_{\text{S1-S1}}\right)^2=& - D_\mu^2 + \frac{ie}{4} \Sigma_{\mu\nu}F_{\mu\nu} +\left(\slashed{D} \Gamma^0\mathcal{H}^{\lambda,g}_{\text{LB,S1-S1}}\,+\,\Gamma^0\mathcal{H}^{\lambda,g}_{\text{LB,S1-S1}}\slashed{D}\right) +(\Gamma^0\mathcal{H}^{\lambda,g}_{\text{LB,S1-S1}})^2,
\end{aligned}
\end{equation}
where $\Sigma_{\mu\nu}= [\Gamma_\mu,\Gamma_\nu]$. Keeping only the first term in the exponent and expanding the rest in series we get
\begin{equation}
\begin{aligned}
&\Tr\left[\bra{x}\,\Gamma^5 e^{-\frac{\left(\slashed{D}^{\lambda,g}_{\text{S1-S1}}\right)^2}{M^2}}\ket{x}\right]\\&=
\Tr\bigg[\bra{x}\,\Gamma^5 e^{\frac{D_\mu^2}{M^2}}\,\,\sum^\infty_{m=0}\frac{1}{M^{2m} m!}\left(-\frac{ie}{4} \Sigma_{\mu\nu}F_{\mu\nu} -\left(\slashed{D} \Gamma^0\mathcal{H}^{\lambda,g}_{\text{LB,S1-S1}}\,+\,\Gamma^0\mathcal{H}^{\lambda,g}_{\text{LB,S1-S1}}\slashed{D}\right) -(\Gamma^0\mathcal{H}^{\lambda,g}_{\text{LB,S1-S1}})^2\right)^m\ket{x}\bigg].
\end{aligned}
\end{equation}
One can verify that the terms with $m=0,1$ vanishes after taking trace with $\Gamma^5$. Only, $m=2$ order terms will give non-zero contribution to the trace in the limit when $M\to \infty$, where we have used $\text{Tr}[\Gamma^5 \Gamma_\mu\Gamma_\nu\Gamma_\rho\Gamma_\lambda]=-8\varepsilon^{\mu\nu\rho\lambda}$ , which are 
\begin{equation}
\lim_{M\to\infty} \frac{1}{2} \Tr[\bra{x}\Gamma^5e^{\frac{D_\mu^2}{M^2}}\left(\frac{ie}{4 M^2} \Sigma_{\mu\nu}F_{\mu\nu}\right)^2\ket{x}]= \frac{e^2}{16\pi^2} \varepsilon^{\mu\nu\rho\lambda}F^{\mu\nu}F^{\rho\lambda}=
-\frac{e^2}{2\pi^2}\vec{E}\cdot\vec{B}~,\quad  B^i= -\frac{1}{2}\varepsilon^{i j k} F^{j k}.
\end{equation} 
and
\begin{equation}
2\times \frac{1}{2} \Tr[\bra{x}\Gamma^5e^{\frac{D_\mu^2}{M^2}}\,\,\left(\frac{ie}{4M^2} \Sigma_{\mu\nu}F_{\mu\nu}\right) \frac{1}{M^2}\left(\slashed{D} \Gamma^0\mathcal{H}^{\lambda,g}_{\text{LB,S1-S1}}\,+\,\Gamma^0\mathcal{H}^{\lambda,g}_{\text{LB,S1-S1}}\slashed{D}\right)\ket{x}].
\label{eqn:AbelianNonAbelianMix}
\end{equation} 
Using the definition of $\Gamma^0\mathcal{H}^{\lambda,g}_{\text{LB,S1-S1}}$ in terms of the non-Abelian fields given in Eq~\eqref{Action_Elegant} the above term in Eq~\eqref{eqn:AbelianNonAbelianMix} in the $M\to \infty$ limit can be 
written as,

\begin{align}
\lim_{M\to\infty} \Tr[\bra{x}\Gamma^5e^{\frac{D_\mu^2}{M^2}}\,\,\left(\frac{e}{4M^4} \Sigma_{\mu\nu}F_{\mu\nu}\right)\Gamma_\alpha\Gamma_i \left[D_\alpha, \mathcal{A}_i^0(g)\right]\ket{x}]~,~i=1,2,3.
 \end{align}

If we now use the definition of the non-Abelian fields in Eq~\eqref{non-Abelian fields} but in the position space,  we  can simplify the logarithm of the Jacobian in Eq~\eqref{eqn:Jacobian} to the following final expression in a compact form,
\begin{equation}
\ln \mathcal{J} =-i\,\int\,d^4x\,\, \alpha(x)\bigg[\frac{e^2}{2\pi^2} \vec{E}\cdot\vec{B}\,\, -\frac{3 e^2\,g}{4\pi^2}\vec{E}\cdot\vec{B}\bigg].
\end{equation}
The Jacobian will also contribute in changing the action due to chiral transformation, apart from the contribution due to the derivative of the chiral current at tree-level, resulting in following anomaly equation,

\begin{equation}\label{anomaly_s1s1}
\partial_\mu J^{5\mu} - \frac{(2-3g)e^2}{2\pi^2} \vec{E}\cdot\vec{B}=0.
\end{equation}
\subsection*{Case 2: S1-2W System}
The derivation of the anomaly equation for the case of two inequivalent nodes follows the same procedure described above except for the point that the modified Dirac operator in this case $\slashed{D}^{\lambda,g}_{\text{S1-2W}}$ is not a Hermitian operator as the off diagonal blocks differ in structure. However, we do need a Hermitian operator so that we can expand the fermionic variables in terms of the eigen spinor of that Hermitian operator as done in the previous section. This problem is resolved if we use the eigen spinors of the Laplacian operators $(\slashed{D}^{\lambda,g}_{\text{S1-2W}})^\dagger\,(\slashed{D}^{\lambda,g}_{\text{S1-2W}})$ and $(\slashed{D}^{\lambda,g}_{\text{S1-2W}})\,(\slashed{D}^{\lambda,g}_{\text{S1-2W}})^\dagger$. Let the eigen spinors of $(\slashed{D}^{\lambda,g}_{\text{S1-2W}})^\dagger\,(\slashed{D}^{\lambda,g}_{\text{S1-2W}})$ is $\Phi_n$ and that of $(\slashed{D}^{\lambda,g}_{\text{S1-2W}})\,(\slashed{D}^{\lambda,g}_{\text{S1-2W}})^\dagger$ is $\xi_n$. Thus
\begin{align}
  &  (\slashed{D}^{\lambda,g}_{\text{S1-2W}})^\dagger\,(\slashed{D}^{\lambda,g}_{\text{S1-2W}}) \Phi_n=\eta_n^2 \Phi_n\\
  &(\slashed{D}^{\lambda,g}_{\text{S1-2W}})\,(\slashed{D}^{\lambda,g}_{\text{S1-2W}})^\dagger \xi_n= \eta_n^2 \xi_n,
\end{align}
where $\eta_n^2$ is the eigen value of the Laplacian operators. Then the fermionic variables can be expanded in terms of these eigen basis following the prescription given in \cite{Fujikawa:2004cx, PhysRevResearch.4.L042004}
\begin{align}
    \psi= \sum_m C_m \Phi_m(x),\,\, \bar{\psi}= \sum_m \bar{C}_m\xi^\dagger_m.
\end{align}
Using this expansion one can show that the regularized Jacobian in this case looks like
\begin{equation}
\begin{aligned}
\ln \mathcal{J} = \lim_{M\to\infty}\,i\,\int\,d^4x\,\, \alpha(x)\,\Tr\left[\bra{x}\,\Gamma^5 \left(e^{-\frac{(\slashed{D}^{\lambda,g}_{\text{S1-2W}})^\dagger\,(\slashed{D}^{\lambda,g}_{\text{S1-2W}})}{M^2}}+ e^{-\frac{(\slashed{D}^{\lambda,g}_{\text{S1-2W}})\,(\slashed{D}^{\lambda,g}_{\text{S1-2W}})^\dagger}{M^2}}\right)\ket{x}\right].
\end{aligned}
\end{equation}

With the modified Dirac operator in Eq~\eqref{eff action in equiv nodes} and performing the trace calculation one obtains the following anomaly equation
\begin{equation}\label{anomaly_s12w}
\partial_\mu J^{5\mu}-\left( 2-\frac{3g}{2}\right)\frac{e^2}{2\pi^2} \vec{E}\cdot \vec{B} =0.
\end{equation}

\section{Heuristic understanding of anomaly via Landau levels}\label{Appendix G}

\begin{figure}[h]
  \centering
  \subfloat[Weyl]{\includegraphics[width=0.45\textwidth]{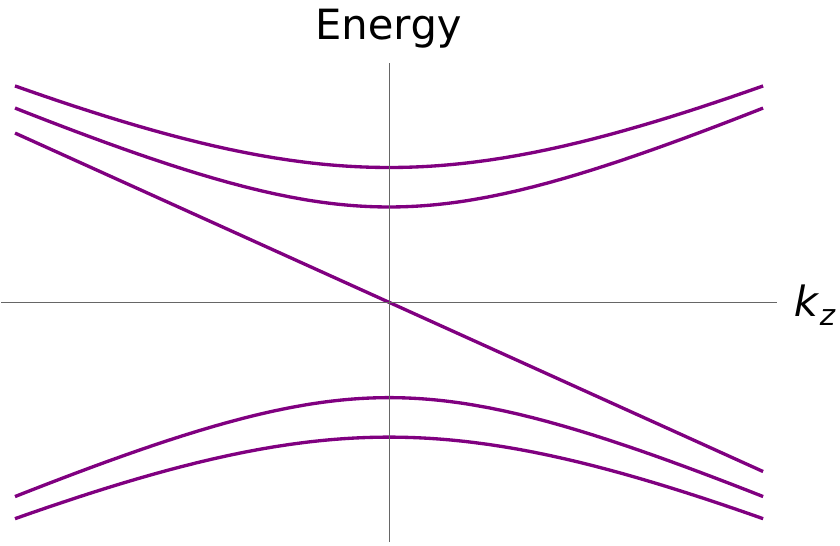}}\hfill
  \subfloat[Double Weyl]{\includegraphics[width=0.45\textwidth]{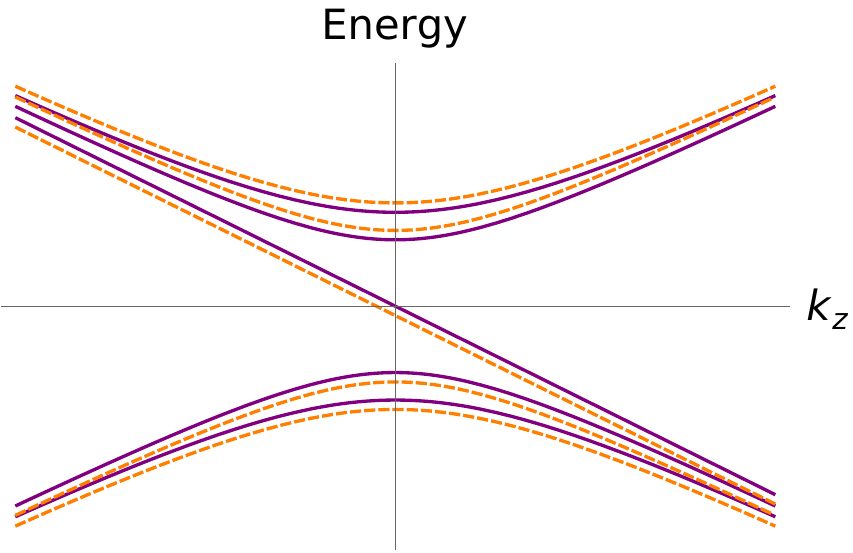}}
  \caption{A few low-energy Landau levels for (a) conventional Weyl and (b) 2-flavor Weyl fermions. The figure illustrates the fact that the number of chiral Landau levels crossing the Fermi level at zero energy is one in (a) and two in (b), which are equal to their respective flavor numbers. In each case, we show only one chiral node. The node with the opposite chirality (not shown) is identical except for the chiral mode(s) having equal but opposite slope.}
  \label{fig:1}
\end{figure}

For the Weyl fermions, which preserve the emergent Lorentz symmetry, the phenomenon of chiral anomaly is understood in terms of charge pumping between left and right chiral nodes via the lowest Landau level (LLL) which is chiral in nature. We would like to discuss this method of understanding chiral anomaly and it's ineffectiveness for Lorentz symmetry broken systems like spin-$1$ fermions.

We first consider Fig.~\ref{fig:1}(a) which depicts the Landau levels formed by Weyl fermions in a constant magnetic field. The LLL, represented by the single chiral band present in Fig.~\ref{fig:1}(a), contributes to the charge pumping phenomenon. This is because the higher bands being hyperbolic in nature contains both the left and right movers and thus does not contribute to any net flow of chiral charge. This leads to a dimensional reduction-- the imbalance in chiral charge for any given chiral node can be described in terms of the chiral LLL which is equivalent to discussing chiral anomaly in 1+1 dimensional model of Weyl fermions. To calculate the rate of creation of the chiral particles we also need to take into account the density of states for these modes. For a system of length $L$ along the direction of the magnetic field the density of states per unit area perpendicular to the magnetic field is given by 
\begin{eqnarray}
   D(\varepsilon)= \frac{L e B}{4\pi^2}.
\end{eqnarray}
Then the rate of change of number density (per unit length) of left (right) handed Weyl fermions is given by~\cite{NIELSEN1983389}
\begin{equation}\label{flow of Fermi level}
    \dot{N}_{L(R)} = \frac{1}{L} \frac{L e B}{4\pi^2} \dot{\varepsilon}_F= -(+)\frac{e B}{4\pi^2} \dot{k}_{zF},
\end{equation}
where we have used the dispersion relation for left (right) handed Weyl fermions at LLL i.e. $\varepsilon= -(+) k_z$. In presence of electric field $E$ along the z-direction the Fermi momentum will flow as $\dot{k}_{zF}= e E$. Using this we have
\begin{equation}
    \dot{N}_{L(R)} = \frac{1}{L} \frac{L e B}{4\pi^2} \dot{\varepsilon}_F= -(+)\frac{e^2 EB}{4\pi^2},
\end{equation}
Thus one get the total imbalance in chiral charge density
\begin{eqnarray}
   \frac{d Q_5}{dt}= \dot{N}_R- \dot{N}_L = \frac{e^2}{2\pi^2} E\, B.
\end{eqnarray}

In case of 2-flavor Weyl fermions, shown in Fig.\ref{fig:1}(b), there are two copies of the chiral Landau levels. Thus, in this case, we simply get 
\begin{eqnarray}
   \frac{d Q_5}{dt}= \dot{N}_R- \dot{N}_L =2 \frac{e^2}{2\pi^2} E\, B.
\end{eqnarray}

\begin{figure}[h]
  \centering
  \subfloat[]{\includegraphics[width=1\textwidth]{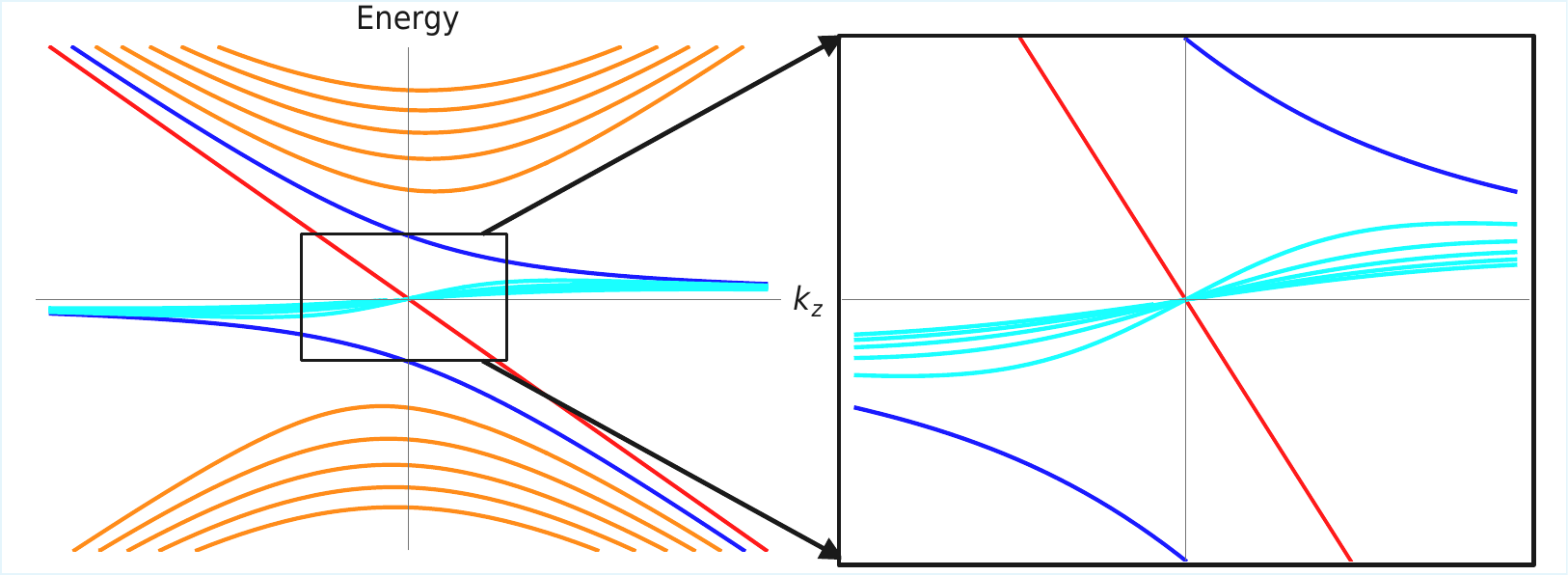}}\hfill
  \subfloat[$g=1$]{\includegraphics[width=0.3\textwidth]{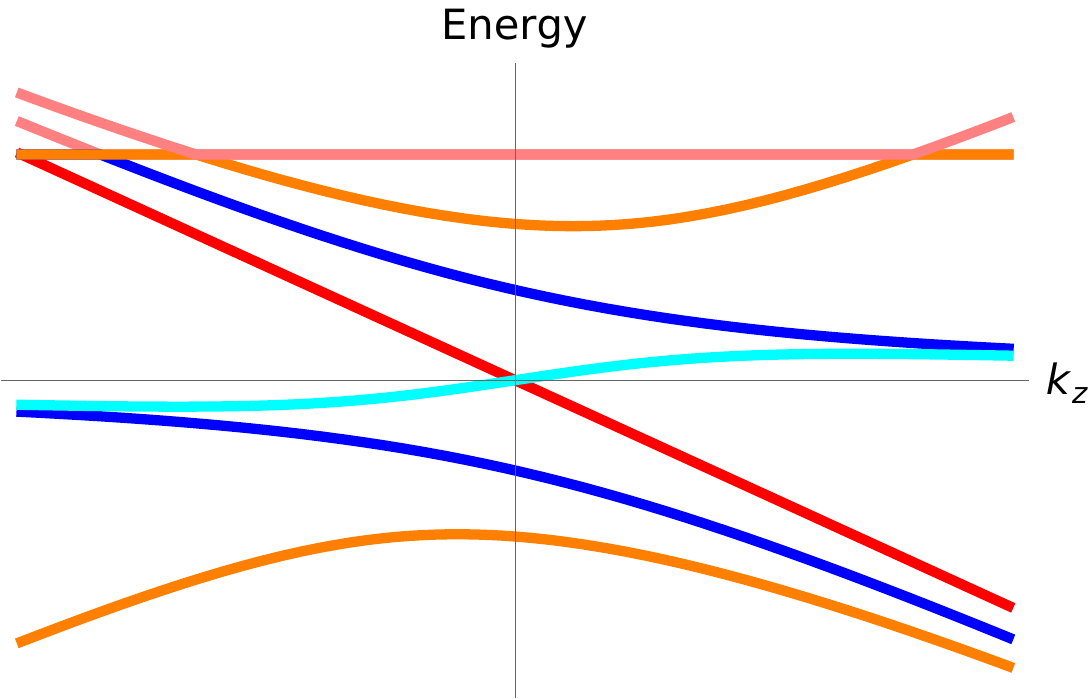}}\hfill
  \subfloat[$g=0.5$]{\includegraphics[width=0.3\textwidth]{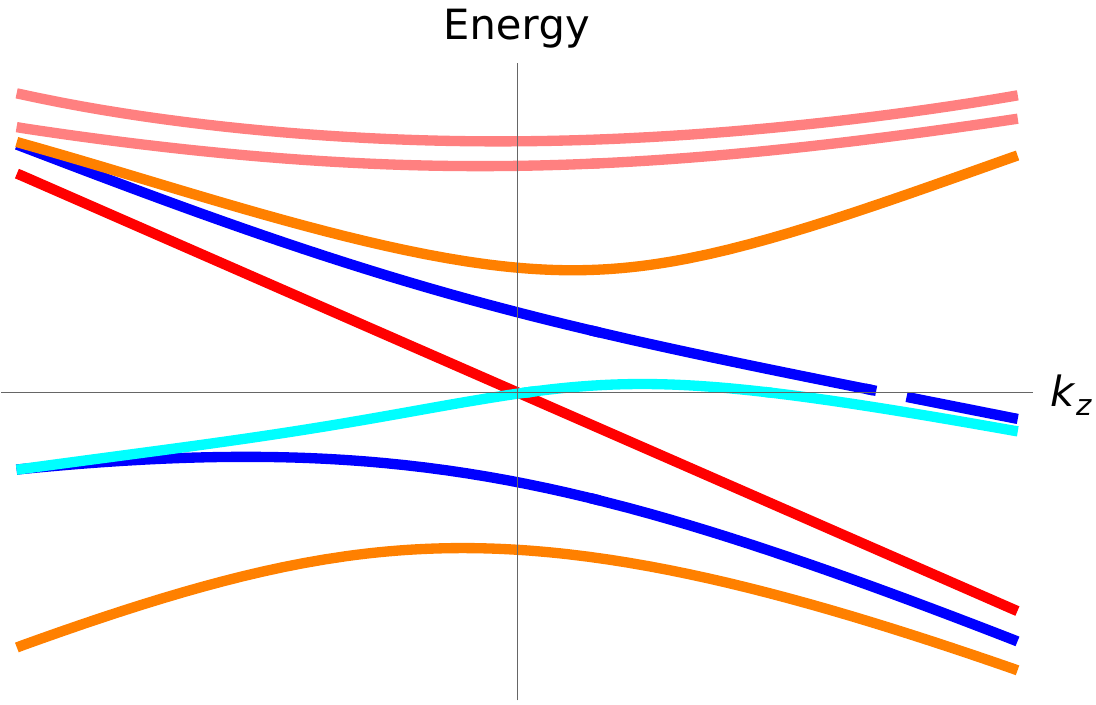}}\hfill
  \subfloat[$g=0$]{\includegraphics[width=0.3\textwidth]{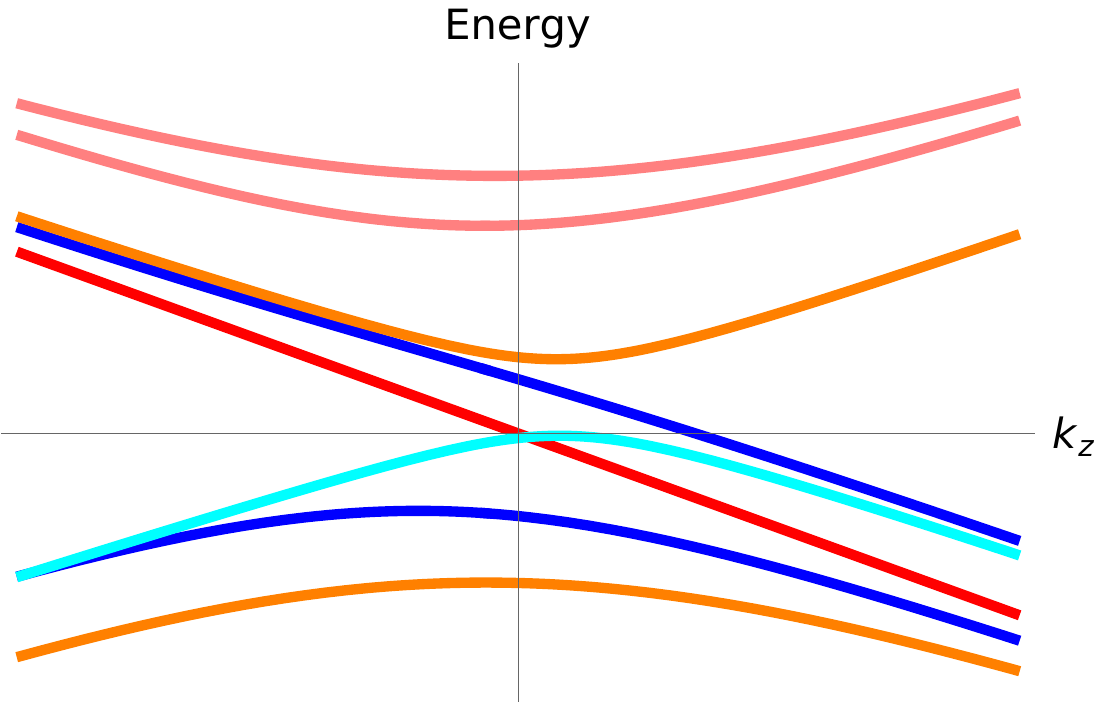}}
  
  \caption{(a) Landau levels (LLs) for spin-1 fermions at $g=1$. The low-energy levels can be clubbed into three kinds: one chiral mode (shown in \textcolor{red}{red}), similar to the Weyl case, crossing zero energy; two other bands (shown in \textcolor{blue}{blue}) meeting the Fermi level asymptotically at large $k_z$, sharing the same sign of slope as the chiral mode, though their slope is energy-dependent; and an infinite number of bands (we show only a few here in \textcolor{cyan}{cyan}) with slope of opposite sign crossing the zero energy. (b-d) show the evolution of the low-energy Landau levels as $g$ is changed from $1$ to $0$. The Weyl-like chiral mode in \textcolor{red}{red} remains unaffected, but the two other kinds of levels deform on changing $g$ (for clarity, we have shown only one level in \textcolor{cyan}{cyan}).}
  \label{fig:2}
\end{figure}

Next, we consider the Landau level structure of spin-$1$ fermions, which is depicted in Fig.~\ref{fig:2}. The Landau levels corresponding to Eq.~(\ref{spin-1 matrices}) can be calculated exactly \cite{bradlyn2016beyond}. We show in Fig.~\ref{fig:2}(a) a few of the low-energy Landau levels. The structure is significantly different from that of Weyl fermions. The striking departure is that, in spite of having a topological charge of $2$, similar to that of 2-flavor Weyl fermions, we do not find two chiral Landau levels crossing the Fermi level at zero energy. Instead we get an \emph{infinite} number of levels crossing zero energy. These can be clubbed into three kinds: one chiral mode (shown in \textcolor{red}{red}), similar to the Weyl case, crossing zero energy; two other bands (shown in \textcolor{blue}{blue}) meeting the Fermi level asymptotically at large $k_z$, sharing the same sign of slope as the chiral mode, though their slope is energy-dependent; and an infinite number of bands (we show only a few here in \textcolor{cyan}{cyan}) with slope of opposite sign crossing the zero energy. The net flow of charge out of the spin-1 node, therefore, is no longer expected to be equal and as simple as the 2-flavor Weyl case, even though both share the same topological charge. \\

The situation becomes even more interesting when we consider the generalized $g-$dependent spin-1 Hamiltonian in Eq.~\ref{effective Spin $1$}. Unlike in the previous scenario (corresponding to $g=1$) where the Landau levels could be calculated exactly, in the general case, this is no longer possible. However, one can use a perturbative treatment in $g-1$. Keeping only leading order contribution in $g-1$, we can construct approximate solutions for the Landau levels. These are depicted in Fig.~\ref{fig:2}(b-d) for various values of $g$. We only show a few representative levels of each category. These bands depend on the value of the parameter $g$ and their chiral nature changes once we vary $g$ from $1$ to $0$ as shown in Figs.~\ref{fig:2}(b-d). We note from Fig.~\ref{fig:2}(b-d) that the chiral band (shown in \textcolor{red}{red}) does not depend on $g$ and remains chiral for all values of $g$. However, this is not the case for the other bands mentioned previously. In particular, the two asymptotic bands, shown in \textcolor{blue}{blue}, gradually becomes chiral as $g$ varies from $1\rightarrow 0$ in Fig.-\ref{fig:2} (b-d). Also, the set of infinite number of bands in Fig.~\ref{fig:2}(a),(b), shown in \textcolor{cyan}{cyan}, which shows chiral features near $k_z=0$, deforms to non-chiral bands as $g$ varies from $1\rightarrow 0$, intersecting twice at zero energy with unequal velocities, as shown in Fig.-\ref{fig:2} (b-d). One can therefore argue that flow of chiral charge in the present scenario would now be dependent on the parameter $g$. Indeed, one can think of modeling these bands around the point $k^i_{z0} (g,B,\lambda)$ where it touches the $k_z$ axis by taking a linear approximation around that point, i.e., in general for any such bands with index $i$ the dispersion near to the touching point would be approximately $|E^i|= v^{\prime i}(g,B,\lambda) (k_z-k^i_{z0} (g,B,\lambda))$. Then Eq.~\eqref{flow of Fermi level} suggests that the velocity $v^{\prime i}(g,B,\lambda)$, given by the slope of the curve at Fermi energy, will appear in the anomaly relation and thus it will become $g$ dependent. This line of argument suggests that in a Lorentz symmetry-breaking scenario, such $g$ dependent terms will indeed appear in the anomaly equation. However, there exist an infinite number of such levels, as discussed previously, making it formidable, if not impossible, to calculate the anomaly equation exactly with this very simple approach.


\section{Comparison with multi-Weyl semimetals}\label{Appendix H}
Our central result---Eqs.~\eqref{anomaly_s1s1} and~\eqref{anomaly_s12w}---demonstrates that, in the absence of Lorentz symmetry, the anomaly coefficient is generally no longer quantized. At this stage, it is instructive to compare our findings with earlier results from a different class of Lorentz-violating systems, namely multi-Weyl semimetals~\cite{PhysRevB.94.195144,PhysRevResearch.2.013007,PhysRevB.96.085201}. Interestingly, despite the lack of Lorentz symmetry, these systems retain a quantized anomaly coefficient. The low-energy Hamiltonian for a given node in a multi-Weyl semimetal, with chirality sign \(s\), is given by
\begin{equation}
    H^{mW}_{s}= s  [k_z\sigma^3 + k^n_{+}\sigma_{-} + k^n_{-}\sigma_{+}],\,  k_{\pm}= \frac{k_x\pm k_y}{\sqrt{2}},\,\,\sigma_{\pm}= \frac{\sigma^1\pm \sigma^2}{\sqrt{2}},\,\,\,,\,\, n=1,\,2,\,3.
\end{equation}
For $n\, >\, 1$ dispersion in the $x-y$ plane becomes non-linear in momentum, making the system manifestly Lorentz symmetry violating. Following Ref.~\cite{PhysRevResearch.2.013007}, a comparison between the spin-1 model and the multi-Weyl model is given below:
\begin{align}
 &  \text{Spin-1 System:} ~\mathcal{A}^0_i= -g\frac{k_i}{2},\, \mathcal{A}^i_0= g\frac{k_i}{2},\,\mathcal{A}^0_0= \frac{\lambda}{4},\, \mathcal{A}^i_i= -\frac{\lambda}{4},\, \mathcal{A}^j_i=0~,~i,j=1,2,3.\label{nA_S1}\\
 & \text{Multi-Weyl System:}\, \mathcal{A}^a_\mu = i\Delta \left( \delta_\mu^x \delta^{a x} + \delta_\mu^y \delta^{a y}\right),\label{nA_multi_Weyl},
\end{align}
where $\Delta$ is a constant, in terms of which both these systems admit a common Hamiltonian described by 
\begin{equation}
    H= H_{2W} + h^{\lambda, g}_{LB},\quad h^{\lambda, g}_{LB}= \mathcal{A}_\mu^a \sigma_\mu \otimes s_a.
\end{equation}
The anomaly equation in both cases can be expressed in the following general way:
\begin{equation}
    \partial_\mu J^5_\mu = \frac{2 e^2}{2\pi^2} \vec{E}\cdot \vec{B} - \frac{e}{16\pi^2} \left(\text{Tr}\left[\left[iD_0, h^{\lambda, g}_{LB}\right] 2B^i \Sigma^i\right]+ \text{Tr}\left[\left[iD^i, h^{\lambda, g}_{LB}\right] \varepsilon^{ijk}\, 2E^j \Sigma^k \right] \right).
\end{equation}
We note that, when the non-Abelian  potential is a constant, independent of momentum, the commutator $\left[iD^\mu, h^{\lambda, g}_{LB}\right]$ vanishes making the second term in right hand side of the above equation zero, which leads to an anomaly coefficient that is quantized by the same value as the topological charge, i.e.,
\begin{equation}
     \partial_\mu J^5_\mu = \frac{2 e^2}{2\pi^2} \vec{E}\cdot \vec{B}.
\end{equation}
This is exactly what happens for the multi-Weyl system represented by Eq~\eqref{nA_multi_Weyl}. On the contrary, in spin-1 systems, the non-Abelian field components are momentum dependent, and the commutator becomes non zero:
\begin{equation}
    \left[iD^\mu, h^{\lambda, g}_{LB}\right]\neq 0.
\end{equation}
This leads to the additional contribution and the anomaly coefficient becomes non-quantized. Note that, even in this case, when we put $g=0$, the momentum-dependence in Eq.~\ref{nA_S1} is lost, and the result reduces to the familiar quantized form. Our final expressions in Eqs.~\eqref{anomaly_s1s1} and~\eqref{anomaly_s12w} reflect this.

\section{Anomaly-induced longitudinal magnetoconductivity}\label{Appendix I}

We begin with the chiral anomaly equation, written explicitly in terms of its temporal and spatial components:
\begin{equation}
\partial_{t}\rho_{5}+\nabla\!\cdot\!\mathbf{j}_{5}
= n\,\frac{e^{2}}{2\pi^{2}}\,
\vec{E}\!\cdot\!\vec{B},
\label{aic1}
\end{equation}
where $\rho_{5}=\rho_{R}-\rho_{L}$ and $\mathbf{j}_{5}=\mathbf{j}_{R}-\mathbf{j}_{L}$ denote the axial charge and current densities, respectively. 
The source term on the right-hand side describes the continuous pumping of charge between opposite chiral sectors. 
In a steady state, this pumping must be balanced by inter-chiral scattering, so that the continuity equation reduces to a competition between generation and relaxation of axial charge:
\begin{equation}
\frac{d\rho_{5}}{dt}
= n\,\frac{e^{2}}{2\pi^{2}}\,
\vec{E}\!\cdot\!\vec{B}
- \frac{\rho_{5}}{\tau_{v}},
\label{aic2}
\end{equation}
where $\tau_{v}$ is the relaxation time between the two sectors. 
Here, we have set $\nabla\!\cdot\!\mathbf{j}_{5}=0$, corresponding to a spatially uniform steady state, which is justified when $\vec{E}$ and $\vec{B}$ are constant. 
Solving Eq.~(\ref{aic2}) in the steady state ($d\rho_{5}/dt=0$) gives
\begin{equation}
\rho_{5}
= n\,\frac{e^{2}\tau_{v}}{2\pi^{2}}\,
\vec{E}\!\cdot\!\vec{B}.
\label{aic3}
\end{equation}
The corresponding axial chemical potential $\mu_{5}=\frac{\mu_R-\mu_L}{2}$ is defined thermodynamically through the axial compressibility $\chi_{5}$:
\begin{equation}
\rho_{5} = \chi_{5}\,\mu_{5}.
\label{aic4}
\end{equation}
A finite $\mu_{5}$ leads to a chiral magnetic effect (CME) current, as shown in Ref.~\cite{Fukushima:2008xe}:
\begin{equation}
\mathbf{j}
 = n\,\frac{e^{2}}{2\pi^{2}}\,
   \mu_{5}\,\vec{B}.
\label{aic5}
\end{equation}
Substituting Eqs.~(\ref{aic3}) and~(\ref{aic4}) yields
\begin{equation}
\mathbf{j}
 =n^{2}\,\frac{e^{4}\tau_{v}}{4\pi^{4}\chi_{5}}\,
  (\vec{E}\!\cdot\!\vec{B})\,\vec{B}.
\label{aic6}
\end{equation}
For $\vec{E}\!\parallel\!\vec{B}$, one obtains an anomaly-induced contribution to the longitudinal magnetoconductivity:
\begin{equation}
\Delta\sigma_{L}
= n^{2}\,\frac{e^{4}\tau_{v}}{4\pi^{4}\chi_{5}}\,
B^{2}.
\label{aic7}
\end{equation}
Equation~(\ref{aic7}) is fully quantum and model-independent: the prefactor is universally fixed by the anomaly coefficient $n$, while $\tau_{v}$ and $\chi_{5}$ are determined by the specific band structure and scattering mechanisms of the system.

Some remarks on the axial compressibility are in order. The relation $\rho_{5}=\chi_{5}\mu_{5}$ in Eq.~(\ref{aic4}) assumes equivalent chiral sectors, as in the standard Weyl and the $S1-S1$ cases, where they have identical dispersions and densities of states for both chirality. 
For asymmetric systems such as the $S1-2W$ model, where the right and left sectors possess different compressibilities $\chi_{R}$ and $\chi_{L}$, the axial charge density takes the more general form
\begin{equation}
\rho_{5} = (\chi_{R}+\chi_{L})\,\mu_{5} + (\chi_{R}-\chi_{L})\,\bar{\mu},
\label{aic8}
\end{equation}
with $\bar{\mu}=(\mu_{R}+\mu_{L})/2$ the mean chemical potential. 
The second term represents a static offset of axial charge that exists even for $\mu_{5}=0$ and does not influence the anomaly--induced transport. 
The effective axial compressibility relevant for magnetotransport is therefore
\begin{equation}
\chi_{5}^{\mathrm{eff}} = \chi_{R}+\chi_{L},
\label{aic9}
\end{equation}
which reduces to $\chi_{5}=2\chi$ in the symmetric limit, where $\chi=\chi_R=\chi_L$. 

Let us now consider various cases. For a standard Weyl system, $n=1$ and $\chi_{5}=\mu^{2}/(\pi^{2}v_{F}^{3})$, reproducing the standard expression of Ref.~\cite{PhysRevB.88.104412} for $\Delta\sigma_L$. 
The divergence of $\Delta\sigma_{L}$ as $\mu\!\to\!0$ in this case reflects the vanishing density of states at a Weyl node; in practice, disorder and interactions regularize this behavior, and replace $\mu$ with an appropriate regularizing scale $\Gamma$. In contrast, for a spin--1 fermion the flat band contributes a finite density of states even at charge neutrality. 
A minimal broadening $\Gamma_{f}$ of the flat-band spectral function,
$A(\varepsilon)=\tfrac{1}{\pi}\tfrac{\Gamma_{f}}{\varepsilon^{2}+\Gamma_{f}^{2}}$, gives
\begin{equation}
\chi_{5}\sim \frac{W_{f}}{\pi\Gamma_{f}},
\label{aic10}
\end{equation}
where $W_{f}$ is the flat-band spectral weight. 
Thus, even at $\mu=0$, the system supports a finite $\mu_{5}=\rho_{5}/\chi_{5}$ and consequently a finite magnetoconductivity. 
In the present system, Lorentz symmetry breaking modifies the anomaly coefficient to
\begin{equation}
n_{\mathrm{S1\text{--}S1}}'=2-3g,
\qquad
n_{\mathrm{S1\text{--}2W}}'=2-\tfrac{3}{2}g,
\label{aic11}
\end{equation}
as obtained from the Fujikawa analysis previously. 
Replacing $n\!\to\!n'$ in Eq.~(\ref{aic7}) gives the measurable, non--quantized magnetoconductivity associated with the anomalous chiral anomaly in spin--1 systems. Indeed, it is now a function of $g$, and can even go to zero when $n'$ goes to zero! Note that, a $g-$dependence is also expected in $\chi_5$ (through $W_f$) and $\tau_5$. However, any such dependence is expected to be weak, as long as the resulting  $g-$dependent bandwidth of the (quasi)flat band remains smaller than the scale $\Gamma_f$. The primary $g-$dependence of $\Delta\sigma_L$ then comes from $n'$, as illustrated above.

    \end{appendix}
\end{widetext}

\bibliography{ref.bib}

\end{document}